%% file: main.tex
\documentclass[sigconf, nonacm]{acmart}

\newcommand\vldbdoi{XX.XX/XXX.XX}
\newcommand\vldbpages{XXX-XXX}
\newcommand\vldbvolume{19}
\newcommand\vldbissue{1}
\newcommand\vldbyear{2026}
\newcommand\vldbauthors{\authors}
\newcommand\vldbtitle{\shorttitle}
\newcommand\vldbavailabilityurl{https://github.com/bytezzz/CAKE}
\newcommand\vldbpagestyle{plain}
\usepackage{tabularx}
\usepackage{multirow}
\usepackage{makecell}
\usepackage{float}
\usepackage{subcaption}
\usepackage[table]{xcolor}
\usepackage{booktabs}
\usepackage{diagbox}
\usepackage{mathtools}
\usepackage{listings}
\usepackage{cleveref}
\usepackage{siunitx}
\usepackage{pifont}
\usepackage{bm}
\usepackage[linesnumbered,ruled,vlined]{algorithm2e}
\DontPrintSemicolon
\SetKwInput{KwIn}{Input}
\SetKwInput{KwOut}{Output}
\SetKwComment{tcp}{// }{}
\SetCommentSty{textit}
\newtheorem{theorem}{Theorem}
\usepackage{enumitem}
\usepackage[utf8]{inputenc}
\DeclareUnicodeCharacter{03B2}{\ensuremath{\beta}}
\setlist[itemize]{
    topsep=4pt,
    itemsep=1pt,
    leftmargin=2em
}
\setlist[enumerate]{
    topsep=4pt,
    itemsep=1pt,
    leftmargin=2.8em
}

\DeclareMathOperator*{\argmin}{arg\,min}

\begin{document}

\title{Piece of \tool{}: Adaptive Execution Engines via Microsecond-Scale Learning}
\include{macros}
\author{Zijie Zhao}
\orcid{0009-0002-2376-6300}
\affiliation{%
  \institution{University of Pennsylvania}
}
\email{bytez@seas.upenn.edu}

\author{Ryan Marcus}
\orcid{0000-0002-1279-1124}
\affiliation{%
  \institution{University of Pennsylvania}
}
\email{rcmarcus@seas.upenn.edu}

\begin{abstract}
Low-level database operators often admit multiple physical implementations (``kernels'') that are semantically equivalent but have vastly different performance characteristics depending on the input data distribution. Existing database systems typically rely on static heuristics or worst-case optimal defaults to select these kernels, often missing significant performance opportunities. In this work, we propose \tool{} (Counterfactual Adaptive Kernel Execution), a system that learns to select the optimal kernel for each data ``morsel'' using a microsecond-scale contextual multi-armed bandit. \tool{} circumvents the high latency of traditional reinforcement learning by exploiting the cheapness of counterfactuals -- selectively running multiple kernels to obtain full feedback -- and compiling policies into low-latency regret trees. Experimentally, we show that CAKE can reduce end-to-end workload latency by up to 2× compared to state-of-the-art static heuristics.
\end{abstract}

\maketitle

\pagestyle{\vldbpagestyle}
\begingroup\small\noindent\raggedright\textbf{PVLDB Reference Format:}\\
\vldbauthors. \vldbtitle. PVLDB, \vldbvolume(\vldbissue): \vldbpages, \vldbyear.\\
\href{https://doi.org/\vldbdoi}{doi:\vldbdoi}
\endgroup
\begingroup
\renewcommand\thefootnote{}\footnote{\noindent
This work is licensed under the Creative Commons BY-NC-ND 4.0 International License. Visit \url{https://creativecommons.org/licenses/by-nc-nd/4.0/} to view a copy of this license. For any use beyond those covered by this license, obtain permission by emailing \href{mailto:info@vldb.org}{info@vldb.org}. Copyright is held by the owner/author(s). Publication rights licensed to the VLDB Endowment. \\
\raggedright Proceedings of the VLDB Endowment, Vol. \vldbvolume, No. \vldbissue\ %
ISSN 2150-8097. \\
\href{https://doi.org/\vldbdoi}{doi:\vldbdoi} \\
}\addtocounter{footnote}{-1}\endgroup

\ifdefempty{\vldbavailabilityurl}{}{
\vspace{.3cm}
\begingroup\small\noindent\raggedright\textbf{PVLDB Artifact Availability:}\\
The source code, data, and/or other artifacts have been made available at \url{\vldbavailabilityurl}.
\endgroup
}

\input{introduction}
\input{sysmodel}
\input{technique}
\input{experiments}
\input{relatedwork}
\input{conclusion}

\begin{acks}
 This work was supported by a gift from Merly AI and the Google ML and Systems Junior Faculty Award.
\end{acks}

\bibliographystyle{ACM-Reference-Format}
\bibliography{ryan-cites-long, references}

\end{document}

%% file: macros.tex
\newcommand{\z}[1]{{\color{purple}[Zijie: #1]}}
\newcommand{\note}[1]{{\color{gray}[Note: #1]}}
\newcommand{\todo}[1]{{\color{red}\textbf{[TODO: #1]}}}
\newcommand{\tool}{\textsc{CAKE}}

\newcommand{\bootstrap}{$\text{\tool{}}^\text{bootstrap}$}
\newcommand{\clt}{$\text{\tool{}}^\text{CLT}$}

\newcommand{\circleOne}{{\ding{182}\xspace}}
\newcommand{\circleTwo}{\ding{183}\xspace}
\newcommand{\circleThree}{\ding{184}\xspace}
\newcommand{\circleFour}{\ding{185}\xspace}
\newcommand{\circleFive}{\ding{186}\xspace}
\newcommand{\circleSix}{\ding{187}\xspace}

\newcommand{\sparagraph}[1]{\vspace{1mm}\noindent {\bf #1}}

%% file: introduction.tex
\section{Introduction}

Low-level database operators often admit multiple physical implementations. For example, a sort operator may use either Quicksort or Heapsort to sort small batches of data. To choose among these alternatives, some DBMSes simply select a worst-case optimal default (e.g., PostgreSQL~\cite{url-postgres} always uses Quicksort for in-memory sorting), or rely on simple heuristics (e.g., DataFusion~\cite{arrow_filter_strategy} switches between slice and bit iterators based on density).

Both strategies have clear drawbacks. Using a single implementation ignores the characteristics of the input data and therefore prevents the system from selecting the most appropriate algorithm for a given workload. Conversely, relying on simple heuristics requires manual tuning, potentially from both the developer and the end user on a per-dataset basis, and requires constant maintenance.

Following~\cite{excalibur}, we refer to these low-level implementations as \emph{kernels}. Virtually every kernel can have multiple implementations, each of which may be optimal for a specific type of input data. For example, some kernels may perform efficiently on nearly-sorted data, while others are better suited for sparse data. Since the optimal kernel depends on properties of the input data, the correct choice might vary for each \emph{morsel}~\cite{morsel} of a query.

\begin{figure}
    \centering
    \includegraphics[width=\linewidth]{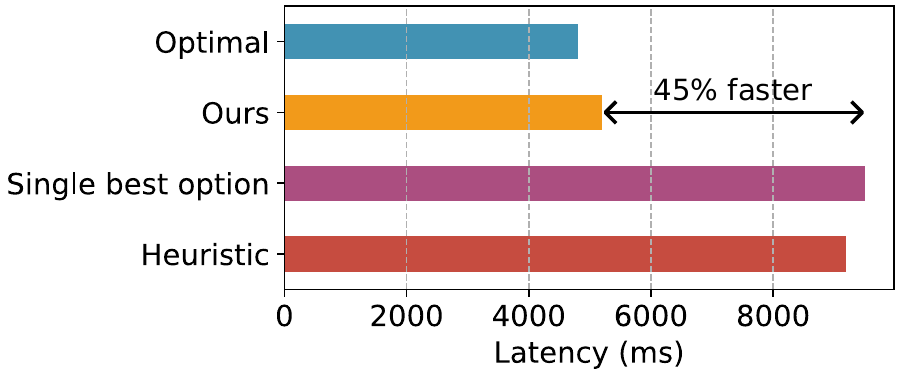}
    \caption{Latency of our IMDb workload using different kernel selection algorithms. Choosing the right kernels for each morsel can have a large performance impact. See Section~\ref{sec:expr}.}
    \label{fig:page1}
\end{figure}

Choosing the correct kernel is essential for performance. Figure~\ref{fig:page1} illustrates the performance of different kernel selection methods on our IMDb workload (see Section~\ref{sec:expr} for details). Notably, the performance gap between choosing the single best static option and choosing the optimal kernel for each data morsel is nearly $2\times$.

This raises a critical question: how can we build DBMSes that select the correct kernel for each morsel without significantly increasing the burden on DBMS developers and without requiring users to manually tune systems for their own datasets? Ideally, developers should never have to design a heuristic, and DBMSes should function efficiently out-of-the-box.

We propose a new reinforcement learning (RL) technique, specifically a contextual multi-armed bandit technique (CMAB)~\cite{bandit_survey} (slightly different from the standard setting, as we allow the model to pay an additional cost to observe the rewards of non-chosen arms), to automatically learn a mapping between morsels and optimal kernels in an online fashion. Our system, \tool{} (\underline{C}ounterfactual 
\underline{A}daptive \underline{K}ernel \underline{E}xecution), tackles this problem using a simple but powerful algorithm based on conditional sampling. Unlike prior work on learned query optimization (e.g.,~\cite{bao,fastgres,hero_qo}), \tool{} operates at operator granularity, making different decisions for each morsel of data in a query. In order to apply RL to a problem like kernel selection, we faced two key challenges.

\sparagraph{Challenge 1: Overhead} Most kernel invocations take dozens to hundreds of microseconds. That means an effective RL agent to select kernels has to operate at microsecond scale; every microsecond of inference time eats into the gains from good kernel selection. This rules out the vast majority of RL techniques, such as those using neural networks~\cite{dqn,ppo,deep_cmab}.  \tool{} resolves this issue with a microsecond-scale contextual bandit that uses locality-weighted resampling to estimate per-kernel latency distributions and certify the optimal choice via empirical confidence tests.

\sparagraph{Challenge 2: Exploration / Exploitation \& Sample Efficiency} All RL techniques must balance exploring new strategies with exploiting existing knowledge~\cite{rl_book}. A central challenge in reinforcement learning is the lack of counterfactuals: once an action is taken and a reward is observed, we never learn what reward would have resulted from a different choice. 
This is not the case in our scenario. Because a single query is composed of thousands of morsels and each kernel invocation typically completes within a few microseconds, counterfactual evaluations can often be obtained at low cost. For example, if two kernel implementations for an operator require \qty{12}{\micro\second} and \qty{6}{\micro\second}, respectively, executing both yields complete feedback at a total cost of only \qty{18}{\micro\second}. Since end-to-end query execution times are on the order of milliseconds, selectively computing such counterfactuals introduces negligible overhead. While exploration is still needed, \tool{} can strategically decide when to collect counterfactual data, allowing \tool{} to be significantly more sample efficient than other approaches.  \tool{} is so sample efficient that \tool{} can learn an effective kernel-selection strategy even \emph{within the execution of just a few queries.}

\vspace{2mm}
We designed \tool{} to be easily integrated into high-performance DBMSes. First,  our developer interface (presented in Section~\ref{sec:interface}) only requires a database engineer to define each kernel implementation and provide a feature extraction function. Second, since \tool{} can potentially learn from each morsel processed by the system, \tool{}'s decision models can converge quickly. While these decision models have only a few microseconds of inference latency (normally within \SI{1}{\us}), \tool{} can almost completely eliminate this cost by compiling a converged model into a \textbf{regret tree}, a special type of decision tree that minimizes the regret of each split. \tool{} is able to train this type of model because it collects counterfactual data. Regret trees are small (less than a cache line) structures with an inference path containing only a few branches, typically measuring less than \qty{20}{\nano\second} when cached.

Experimentally, we show that \tool{} learns accurate kernel selection algorithms with low overhead, and can reduce end-to-end workload latency on real datasets by up to 2x. We additionally study \tool{}'s hyperparameters sensitivity, sample efficiency, and regret tree optimizations. 
  
This work makes the following contributions:
\begin{itemize}
    \item{We present \tool{}'s \emph{core algorithm}, designed to work at microsecond scale using empirical or approximated confidence intervals over empirical weighted samples, as well as a proof of the algorithm's convergence.}
    \item{We show how learned models can be effectively compiled into efficient static \emph{regret trees}.}
    \item{We provide an experimental evaluation of \tool{}, showing that \tool{} can reduce total and tail latency by a factor of up to 2x.}
\end{itemize}

The rest of this paper is organized as follows. We present our system model in Section~\ref{sec:sysmod}. We outline the implementation of \tool{}, including a proof of \tool{}'s convergence, in Section~\ref{sec:cake}. We present experimental results in Section~\ref{sec:expr}. Finally, we discuss related works in Section~\ref{sec:rw}, and conclude in Section~\ref{sec:conclusion}.

%% file: sysmodel.tex
\section{System Model}
\label{sec:sysmod}
In this section, we explain the key assumptions made by \tool{}.

\begin{figure*}[htbp!]
    \centering
    \includegraphics[width=0.9\linewidth]{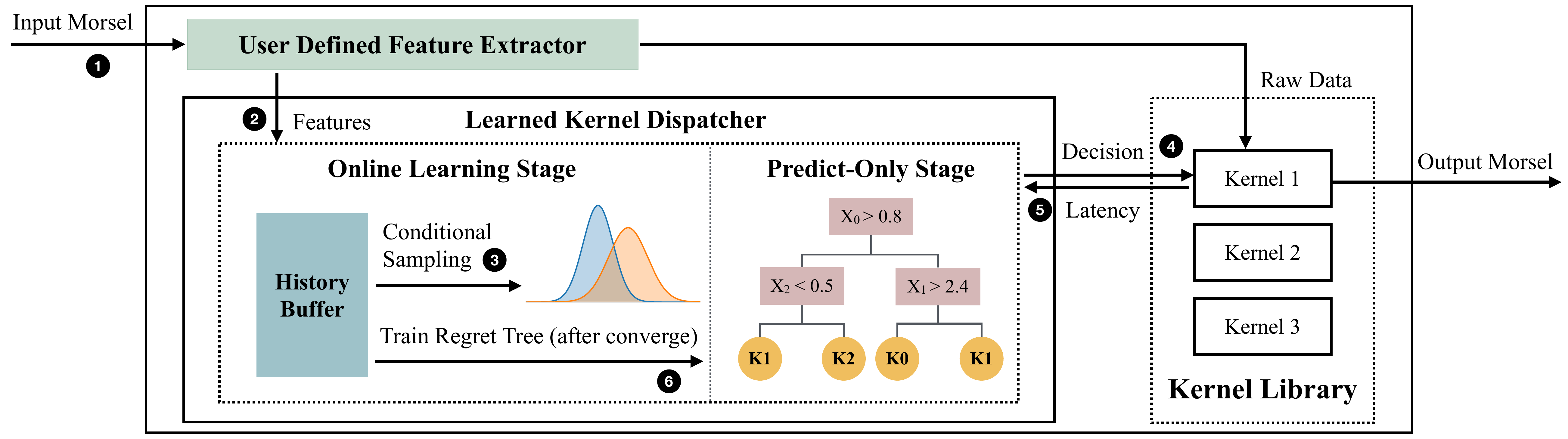}
    \caption{System Diagram}
    \label{fig:diagram}
\end{figure*}

\sparagraph{Morsel-driven parallelism} We assume an execution engine based on morsel-driven parallelism~\cite{morsel}. Morsel-driven parallelism is a query execution framework designed for many-core architectures. A morsel-driven system divides input data into small, constant-sized fragments (e.g., 10,000 tuples for each morsel). Workers, which are pinned to hardware cores, grab a morsel from a queue and run that morsel through the entire pipeline. This execution strategy is cache and register friendly, and enables advanced features like elasticity.

\sparagraph{Kernel selection} Since each pipeline is composed of multiple operators, a morsel-driven database system must choose operator implementations for each operator. Following~\cite{excalibur}, we refer to these implementations as \emph{kernels}. The optimal kernel for a particular operator depends on the data distribution of the morsel. For example, a kernel that is optimal on average can be suboptimal for a substantial fraction of morsels: a vectorized kernel may pay fixed setup costs that only amortize on heavy morsels, while a branchy scalar kernel may excel on light morsels but stall on heavy ones. Some DBMSes use a single kernel for each operator, while others employ simple heuristics for data-dependent kernel selection~\cite{arrow_compute,duckdb}. Unfortunately, these simple heuristics have several drawbacks: (1) they must be manually engineered for each operator, and must be updated when new kernels are developed, and (2) they must be tuned per dataset for optimal performance. Instead of manually developing heuristics, we argue that \textbf{the execution engine should \emph{learn} to select kernels at morsel granularity}, adapting the implementation to each morsel’s local characteristics to minimize end-to-end query time.

\sparagraph{Requirements for learned kernel selection} Learning to choose among semantically equivalent kernels is difficult because the selector must be cheaper and more reliable than the kernels it governs. First, many kernels complete in microsecond time, so \emph{feature extraction and inference must remain at microsecond scale (ideally below \qty{1}{\micro\second}),} otherwise the predictor’s overhead dominates and eliminates any benefit. Second, \emph{training overhead must be negligible}: the system cannot pause query processing for batch retraining, and must instead learn online under a tight CPU budget while continuing to execute morsels. Third, decision quality must be sufficient to \emph{improve end-to-end latency}: the learned policy must be accurate in the regions where latency gaps are significant rather than merely achieving high aggregate classification accuracy -- in other words, the model must be right where it matters. 

Unfortunately, off-the-shelf RL algorithms fail to satisfy these requirements. For example, techniques using expensive deep neural networks to model the policy~\cite{reinforce}, value function~\cite{dqn}, or both~\cite{ppo, rejoin_tail}, could certainly achieve high accuracy, but are far too heavyweight~\cite{thodoroff_benchmarking_2022}. On the other hand, multi-armed bandit techniques~\cite{bandit_survey} are lightweight enough~\cite{excalibur, microadapt}, but make too many assumptions about the relationship between features and rewards~\cite{li_contextual-bandit_2010}.

\sparagraph{Opportunities} We exploit two structural advantages to make learning practical in this setting. First, the system generates abundant data: a single query could be made up of thousands of morsels, meaning that a single query can generate thousands of training instances. This is in contrast to prior work applying ML to query optimization, where feedback is sparse (one per query)~\cite{bao, balsa, roq}. Second, counterfactual feedback is inexpensive and often available: by occasionally running multiple candidate kernels on the same morsel, we can observe both the ``right'' and ``wrong'' outcomes (e.g., \qty{1}{\milli\second}  vs. \qty{2}{\milli\second}) at a total cost comparable to their sum (e.g., \qty{3}{\milli\second}), enabling direct measurement of regret -- something that most RL algorithms painstakingly approximate~\cite{rl_book}. This contrasts sharply with the query optimization settings, where exploring a poor plan can incur minutes-to-hours of wasted time, making counterfactuals effectively unattainable. These two structural advantages also provide guidance for the direction of designing new algorithms.

\subsection{Architecture} 
Our system, \tool{}, can automatically learn which kernels to pick for which morsels. Database developers only need to surface the list of available kernels and compute some features. \tool{}'s architecture is illustrated in the \autoref{fig:diagram}, and described below.

\sparagraph{Feature Extraction}   As an \circleOne~\emph{Input Morsel} arrives, it is passed through a \emph{User Defined Feature Extractor}. This user-defined component extracts a context vector (e.g., selectivity, length) representing the morsel's data distribution, which is passed to the \emph{Learned Kernel Dispatcher}. Developers are responsible for ensuring this feature extraction has trivial overhead -- we show this is easily achievable with techniques such as sampling. If a heuristic already exists, developers can re-use the inputs to that heuristic as features.

\sparagraph{Online Learning Stage} Initially, the dispatcher operates in the \emph{Online Learning Stage} shown on the left of \autoref{fig:diagram}. It maintains a \emph{History Buffer} containing past contexts and their observed latencies.  Given the \circleTwo extracted features, using a carefully designed conditional sampling algorithm, \circleThree \tool{} weighs this historical evidence to decide which kernel to run. It may choose to exploit the best-known kernel or explore by gathering counterfactual data (running alternative kernels). The learned kernel dispatcher will make the decision of which kernel(s) to execute, \circleFour then executes the kernel(s) on the input morsel. The resulting latency \circleFive is fed back into the History Buffer to refine future decisions.

\sparagraph{Predict-Only Stage} Once the learning model converges, \tool{} transitions to the Predict-Only Stage, shown on the right of \autoref{fig:diagram}. The complex probabilistic model is \circleSix compiled into a lightweight \emph{regret tree}. This tree uses simple conditional checks on the input features to route the morsel to the optimal implementation (K1, K2, etc.). This stage eliminates the computational overhead of learning entirely, ensuring negligible latency during steady-state execution.

Together, these stages enable \tool{} to learn online when local evidence is scarce (online learning stage) and run at near-zero overhead once decisions stabilize (predict-only stage), enabling per-morsel kernel selection that improves end-to-end latency without changing the developer-facing execution interface.

%% file: technique.tex
\section{CAKE}
\label{sec:cake}

This section describes the details of \tool{}. Section~\ref{sec:interface} describes the interface that \tool{} gives to database developers, along with an example implementation. Section~\ref{sec:algo} describes the core online learning algorithm, along with a convergence proof and optimizations. Finally, Section ~\ref{sec:regret_tree} explains how \tool{} ``freezes'' converged models into regret trees with nearly zero inference overhead.

\subsection{Interface}
\label{sec:interface}

~\autoref{code:inference} shows a minimal interface to use \tool{}. \autoref{code:inference} uses Python syntax for readability and conciseness; our prototype system is implemented in Rust. Users (DBMS developers) are required to provide a list of candidate kernel implementations $\mathcal K = \{k_1, \dots, k_m\}$, as well as a feature extraction function to extract feature $\bm{q} = \{q_1, \dots, q_n\}$ from the input morsel. Furthermore, to ensure robustness, users may optionally specify a fallback kernel and a timeout threshold. If the execution time of any kernel exceeds this threshold during the online learning stage, \tool{} terminates the learning process and defaults to the fallback kernel for all subsequent executions. Our design goal is to impose no additional usage burden on users, allowing them to seamlessly transition from heuristic-based kernel selection to \tool{} while transparently benefiting from the performance gains enabled by adaptive switching.

\sparagraph{Example} To implement a kernel that extracts certain rows of a morsel using a bitmap and copies the selected entries into an output buffer, developers have implemented two alternative copy strategies in the Arrow ~\cite{noauthor_apache_nodate} Rust library (used by DataFusion~\cite{datafusion}):

\begin{enumerate}
    \item \textbf{Index Iterator}: Enumerate true-bit positions and gather corresponding elements into new contiguous buffers.
    \item \textbf{Slice Iterator}: Collapse the boolean mask into contiguous true runs and, for each, bulk-copy those spans from the source into the output.
\end{enumerate}

Intuitively, when the bitmap is dense and contains long runs of true values, the Slice Iterator is preferable because it reduces the number of expensive \texttt{memcpy} calls. In contrast, when the bitmap is sparse, the Index Iterator is more efficient, as it avoids the overhead of compacting bits. To apply \tool{} to this problem, developers can define a feature extractor that computes the bitmap selectivity (i.e., the fraction of true bits) and provide both iterator implementations; \tool{} then makes the selection dynamically based on runtime information.

The Arrow library contains a hand-crafted heuristic to select between the two based on the sparsity of the bitmap~\cite{arrow_filter_strategy}. Unfortunately, this heuristic only considers the sparsity of the bitmap, and not the density of the runs. Therefore, the Slice Iterator could be selected for bitmaps like $[(110)_n]$ (the pattern 110 repeated $n$ times), but not for bitmaps like $[0_n 1_k]$ when $k < n$. When using the same information as the heuristic, \tool{} learns the relationship between sparsity and the optimal kernel on a per-dataset basis. We demonstrate experimentally that \tool{} learns a significantly better heuristic in Section~\ref{sec:expr}. \tool{} could even be augmented with additional features, like the average run length in a sample, to learn an even more robust policy.

\noindent\begin{minipage}{\linewidth}
\begin{lstlisting}[
language=Python,
basicstyle=\ttfamily\small,
caption={Minimal interface, using Python syntax for readability and conciseness.},
captionpos=b,
label={code:inference},
frame=none]
# Different kernel implementations
def kern_a(x: Morsel, alpha: float, beta: int)
    -> Morsel: pass

def kern_b(x: Morsel, alpha: float, beta: int)
    -> Morsel: pass

class Filter(AdaptiveKernel):
    KERNELS = [kern_a, kern_b]
    def extract_features(cls, x: Morsel):
        return [x.selectivity, x.length]

# Execute adaptive kernel selection
out = Filter.adaptive_execute(data, 0.1, 3)
\end{lstlisting}
\end{minipage}

\subsection{Core Learning Algorithm}
\label{sec:algo}

Having established the developer interface and system architecture, we now turn to the decision-making logic that drives \tool{}. The challenge lies in selecting the optimal kernel for each morsel under tight latency constraints. The following subsections detail our formal model, the learning algorithm, and the theoretical guarantees underpinning our approach.

\subsubsection{Problem Definition.}
Formally, a query execution is decomposed into a sequence of morsels $\{u_t\}_{t=1}^T$.
The developer provides a feature (context) extractor $x(\cdot)$, yielding a context vector
$\bm{q}_u = x(u)$.
Let $\mathcal{K}$ denote the set of candidate kernel implementations. All kernels in $\mathcal{K}$ are semantically
equivalent (they produce the same logical result), but their runtimes may differ.
For a context $\bm{q}$ and kernel $k\in\mathcal{K}$, we model the runtime as a random variable $Y_k(\bm{q})$.

The system observes contexts sequentially, i.e., $\bm{q}_t = x(u_t)$ for $t=1,\dots,T$.
Before making the decision at time $t$, the policy observes a history dataset $\mathcal{D}$ consisting of previously
seen contexts, selected kernels, and observed runtimes.
At time $t$, the policy chooses a \emph{nonempty} sequence of kernels
$(k_{t,1},k_{t,2},\dots,k_{t,m_t})$ with $k_{t,j}\in\mathcal{K}$ and $m_t\ge 1$.
The first kernel $k_{t,1}$ is executed to produce the morsel's output and its runtime is observed.
Any additional kernels $\{k_{t,j}\}_{j\ge 2}$ are optional counterfactual measurements: they incur additional runtime
cost but provide only timing observations (no additional semantic benefit).

The time cost incurred at step $t$ is \[
C_t \;=\; \sum_{j=1}^{m_t} Y_{k_{t,j}}(\bm{q}_t),
\]and the total cost of a policy $\pi$ over the entire query is
\[
C(\pi) \;=\; \sum_{t=1}^T C_t.
\]Our objective is to find a policy $\pi$ that minimizes the expected total cost (i.e., an optimal policy):\[
\pi = \argmin_{\pi}\; \mathbb{E}[C(\pi)]
\quad \text{s.t.} \quad m_t \ge 1,\;\; \forall t\in[T].
\]

\begin{algorithm}[htbp!]
\caption{$\tool{}^{Bootstrap}$}
\label{alg:rbf_bootstrap_ci}
\KwIn{Query feature vector $\bm q$; historical dataset $\mathcal D=\{(\bm x_i,\bm y_i)\}_{i=1}^n$ where $\bm y_i=\{y_{i,k}\}_{k\in\mathcal K}$ are per-kernel latencies (smaller is better); significance level $\alpha$; bandwidth $h$; minimum effective sample size $N_{\min}$; bootstrap replicates $B$.}
\KwOut{Selected kernel $k_{\mathrm{sel}}$.}
\BlankLine
\For{$i \gets 1$ \KwTo $n$}{
  $w_i \gets \exp\!\left(-\frac{\|\bm x_i-\bm q\|_2^2}{h^2}\right)$
}
$\tilde w_i \gets \dfrac{w_i}{\sum_{j=1}^n w_j}$ \tcp*[r]{discrete distribution $P(i)=\tilde w_i$}
$N_{\mathrm{eff}} \gets \dfrac{1}{\sum_{i=1}^n \tilde w_i^2}$ \tcp*[r]{effective sample size}

\If{$N_{\mathrm{eff}} > N_{\min}$}{

  \tcp{Bootstrap sample means under $P(i)=\tilde w_i$}
  \For{$b \gets 1$ \KwTo $B$}{
    Sample counts $\bm c^{(b)} \sim \mathrm{Multinomial}\!\left(n,\tilde{\bm w}\right)$ \;
    \ForEach{$k \in \mathcal K$}{
      $\bar y^{(b)}_{k} \gets \dfrac{1}{n}\sum_{i=1}^n c^{(b)}_i\, y_{i,k}$ \tcp*[r]{sample mean}
    }
  }
  $k^\star \gets \argmin_{k\in\mathcal K}\ \sum_{i=1}^n \tilde w_i\, y_{i,k}$

  $\alpha_{\mathrm{adj}} \gets \alpha / \max(1,|\mathcal K|-1)$ \tcp*[r]{Bonferroni Adjustment}
  $q_{\mathrm{lo}} \gets \alpha_{\mathrm{adj}}/2$\,,
  $\textsc{confident} \gets \textbf{true}$

  \ForEach{$k \in \mathcal K\setminus\{k^\star\}$}{
    $\Delta^{(b)} \gets \bar y^{(b)}_{k} - \bar y^{(b)}_{k^\star}$ \tcp*[r]{$b=1,\dots,B$}
    $L \gets \mathrm{Quantile}(\{\Delta^{(b)}\}_{b=1}^B, q_{\mathrm{lo}})$\;
    \If{$L \le 0$}{
      $\textsc{confident} \gets \textbf{false}$ \tcp*[r]{CI overlaps 0 or favors $k$}
    }
  }

  \If{$\textsc{confident}$}{
    \Return $k^\star$ \tcp*[r]{exploit}
  }
}

\BlankLine
\textsc{Explore}:\;
Run all kernels on the current query to obtain $\bm y=\{y_k\}_{k\in\mathcal K}$\;
$\mathcal D \gets \mathcal D \cup \{(\bm q,\bm y)\}$\;
\end{algorithm}

\subsubsection{Learning through bootstrapping counterfactuals.} The core algorithm of \tool{} is given in Algorithm~\ref{alg:rbf_bootstrap_ci}.
This core algorithm uses locality-weighted evidence and a bootstrap-based test to decide when to exploit a single kernel and when to pay the extra cost to explore (i.e., execute counterfactuals). The underlying intuition is as follows: in the feature space, \textbf{historical observations that are closer to the current morsel's features are more predictive of kernel selections for the current morsel}, and the decision of whether to explore depends on \textbf{whether the historical points provide a consistent assessment of the performance} at the current point. Therefore, if there are not enough samples in the history buffer relevant to the current morsel, we should explore. If the samples relevant to the morsel are ambiguous, we should explore. Otherwise, we exploit.

Next, we describe the algorithm precisely. For the current incoming morsel with context vector $\bm q$, we first identify the most relevant historical executions by assigning each record $(\bm x_i,\bm y_i)\in\mathcal D$ a weight based on its attenuated (through an RBF kernel) distance to $\bm q$ (line 1-2). Then, we normalize these weights to form a sampling distribution over past observations (line 3). We compute the effective sample size~\cite{effective_sample_size} $N_{\mathrm{eff}}$, which summarizes how much independent local support the history provides around $\bm q$; if $N_{\mathrm{eff}}$ is below a threshold $N_{\min}$, the algorithm treats the neighborhood as under-supported and immediately explores (line 5).
We next repeatedly draw bootstrap samples from the local distribution defined above to form multiple sample ``jury'' sets (line 7), and use the mean latencies for each kernel over these samples to approximate the per-kernel latency at the query point (line 9). To determine a consensus, we compute the empirical distribution of the pairwise kernel runtime differences (line 14) and make the final decision based on empirical confidence intervals (line 16-17). If we cannot identify the optimal kernel at the given confidence level, we perform an exploration step by executing all kernels on the current morsel, record the full latency vector, and append it to $\mathcal D$, thereby improving local coverage for future contexts (line 20-22).

This design ensures that \tool{} seamlessly switches between performing low-overhead exploitation when the stored history is decisive, and performing targeted exploration when the local evidence is insufficient or statistically ambiguous.

Next, we give a sketch of the proof of \bootstrap's convergence. For the full proof, see the online appendix~\cite{cake_appendix}.

\begin{theorem}
\label{thm:converge}
\textbf{Selection Consistency}. As the size of observed history $n$ increases, the probability of \tool{} selecting the suboptimal kernel for a query morsel $q$ approaches zero linearly assuming variance-bounded noise, and exponentially assuming sub-Gaussian noise.
\end{theorem}

\sparagraph{Proof Sketch.} The core challenge to selecting kernels is understanding the relationship between an incoming morsel's features and prior observations (which are noisy). To estimate the latency of a kernel $k$ at a specific query point $\bm{q}$, written $E[\bm{q}, k]$, \tool{} uses a weighted average of the historical data $\mathcal{D}$, where weights decay with distance from $\bm{q}$ (the RBF kernel). For brevity, we use a morsel $\bm{a}$
to denote a morsel whose feature (context) vector is $\bm{a}$. Letting $\tilde w(\bm{a}, \bm{b})$ be the normalized weight distance from $\bm{a}$ to $\bm{b}$, and letting $f_k(\bm{a})$ be the true latency (without noise) of executing morsel $\bm{a}$ with kernel $k$, \tool{} computes:

\begin{equation}
E[\bm{q}, k] = \sum_{\bm{d} \in \mathcal{D}} \tilde w(\bm{q}, \bm{d}) \times L(\bm{d}, k)
\end{equation}
where $L(\bm d, k)$ denotes the historically observed latency (with noise) of kernel $k$ when processing morsel $\bm d$. Errors in this estimate come from two sources: \textbf{bias} from averaging the points nearby but not identical to $\bm{q}$, and \textbf{variance} from the noisy measurement of latency.

To prove \autoref{thm:converge}, we need following assumptions:
\begin{enumerate}
    \item Smoothness (Lipschitz): we assume $f_k(\bm{x})$ is smooth. If two morsels $\bm a$ and $\bm b$ are similar, we assume there is an $L$ such that $|f_k(a) - f_k(b)| \leq L ||a - b||$.
    \item Separation (Margin): we assume there is a unique best kernel, $k^*$, at a point $\bm{x}$ that beats the second best by a margin of $\Delta > 0$. This is not necessarily true, but when two kernels are tied, we do not care which one \tool{} picks.
    \item Noise: we assume that execution time measurements have an error, but that the error has mean 0 and a bounded variance $\sigma^2$.
\end{enumerate}

Because we only need to identify the optimal kernel (i.e., \tool{} should make the consistent selection as an oracle selector that relies on the true latency functions $f$), it suffices to show that the total estimation error in the estimated gap $E[\bm{q}, k]-E[\bm{q}, k^\star]$ (i.e., the bias-plus-variance contributions from both $k^\star$ and $k$) is strictly smaller than the true margin $\Delta(\bm q)$, so the gap retains the correct sign and the argmin is preserved. Next, we bound the bias and variance of \tool{}'s estimate.

Note that because \tool{} estimates latency using a weighted average of neighboring morsels, the estimator incorporates samples whose contexts differ from $\bm q$, which introduces bias. Under our smoothness assumption, this bias is governed by the typical distance between the query and points drawn from the local distribution~\cite{cake_appendix}, formally $\mathbb E_{X\sim P_h(\cdot\mid \bm q)}[||X-\bm q||]$, where $P_h$ is the local sampling distribution and $h$ is the bandwidth. This matches the intuition that smoothing over too many far-away points increases the bias. We show in the online appendix~\cite{cake_appendix} that $\mathbb E_{X\sim P_h(\cdot\mid \bm q)}[||X-\bm q||]$ converges to a constant proportional to $h$ as $n$ increases; consequently, the estimator’s bias is $O(h)$. Intuitively, when we restrict attention to a sufficiently small neighborhood, the weighted average of nearby samples provides a good approximation at $\bm q$.

Unfortunately, even if we have data in a small neighborhood around $\bm q$, those measurements still have noise. We rely on the effective sample size, $N_\text{eff}$, which counts how many relevant history points we have found. Since we assumed a bounded variance of $\sigma^2$, the variance of our weighted average is $\frac{\sigma^2}{N_{eff}(\bm q)}$.

Finally, for \tool{} to select the wrong kernel, the noise must be large enough to overcome the true margin $\Delta$ (minus the small geometric bias from neighbors not being exactly like $\bm q$). \begin{equation*}
P(\text{wrong choice}) \approx P(\varepsilon_{i,k}-\varepsilon_{i,k^\star} > \Delta -\text{bias})
\end{equation*}
Using Chebyshev's inequality, we have \begin{equation*}
Prob(\text{wrong choice}) \leq  \frac{4\sigma^2}{(\Delta - bias)^2}\cdot\frac{1}{N_{eff}(\bm q)}
\end{equation*} As the size of the history grows, $N_\text{eff}$ grows linearly (rigorous proof in~\cite{cake_appendix}) , so the probability of \tool{} picking an incorrect kernel decays as the history grows. In the appendix, we show that, by assuming sub-Gaussian noise, this convergence is exponential. \qed

\begin{algorithm}[t]
\caption{$\tool{}^{CLT}$}
\label{alg:rbf_kernel_select}
\KwIn{Query feature vector $\bm q$; historical dataset $\mathcal D=\{(\bm x_i,\bm y_i)\}_{i=1}^n$ where $\bm y_i=\{y_{i,k}\}_{k\in\mathcal K}$ are per-kernel latencies; significance level $\alpha$; bandwidth $h$; minimum effective sample size $N_{\min}$.}
\KwOut{Selected kernel $k_{\mathrm{sel}}$.}
\BlankLine

\textit{Compute $\tilde w_i$ and $N_{\text{eff}}$, identical to \autoref{alg:rbf_bootstrap_ci} line 1-4}\;
\If{$N_{\mathrm{eff}} > N_{\min}$}{

  \ForEach{$k \in \mathcal K$}{
    $\hat\mu_k \gets \sum_{i=1}^n \tilde w_i\, y_{i,k}$\;
    $\hat\sigma_k^2 \gets \dfrac{1}{n}\Big(\sum_{i=1}^n \tilde w_i\, y_{i,k}^2 - \hat\mu_k^2\Big)$
    \tcp*[r]{$\bar y_k \approx \mathcal N(\hat\mu_k,\hat\sigma_k^2)$}
  }

  $k^\star \gets \argmin_{k\in\mathcal K}\ \hat\mu_k$\,,
  $\textsc{confident} \gets \textbf{true}$\;
  $\alpha_{\mathrm{adj}} \gets \alpha / \max(1,|\mathcal K|-1)$ \tcp*[r]{Bonferroni Adjustment}

  \ForEach{$k \in \mathcal K\setminus\{k^\star\}$}{
    $z \gets \dfrac{\hat\mu_k-\hat\mu_{k^\star}}{\sqrt{\hat\sigma_k^2+\hat\sigma_{k^\star}^2}}$\;
    \If{$z \le z_{1-\alpha_{adj}}$}{
      $\textsc{confident} \gets \textbf{false}$\tcp*[r]{fail to reject $\mu_{k^\star}\ge \mu_k$}
    }
  }

  \If{$\textsc{confident}$}{
    \Return $k^\star$ \tcp*[r]{exploit}
  }
}

\BlankLine
\textsc{Explore}: \textit{Same as \autoref{alg:rbf_bootstrap_ci} line 20-22}\;
\end{algorithm}

\sparagraph{Optimization: CLT-based confidence bounds} The previously described algorithm, $\tool{}^{\mathrm{Bootstrap}}$, explicitly approximates the local sampling distribution by drawing weighted bootstrap replicas and constructing empirical confidence bounds on mean-latency gaps. Unfortunately, that sampling procedure can be expensive (we demonstrate this experimentally in Section~\ref{sec:expr}). Here, we propose an optimized algorithm, $\tool{}^{\mathrm{CLT}}$, which replaces the entire resampling stage with a closed-form normal approximation. For each kernel $k$, $\tool{}^{\mathrm{CLT}}$ estimates the local mean latency by the same weighted mean $\hat\mu_k=\sum_i \tilde w_i y_{i,k}$, which in our case is exactly the point estimation method known as Nadaraya-Watson kernel regression~\cite{nadaraya_estimating_1964}. But instead of generating bootstrap sample means $\{\bar y_k^{(b)}\}_{b=1}^B$,  $\tool{}^{\mathrm{CLT}}$ estimates the variance of the (weighted) sample mean via a plug-in second-moment calculation: \begin{equation*}
\hat\sigma_k^2=\frac{1}{n}\left(\sum_i \tilde w_i y_{i,k}^2-\hat\mu_k^2\right)
\end{equation*}

By the central limit theorem (CLT), we can treat the sample mean as approximately Gaussian, $\bar y_k\approx \mathcal N(\hat\mu_k,\hat\sigma_k^2)$. The algorithm then selects the provisional winner $k^\star=\arg\min_k \hat\mu_k$, and certifies it using a parametric z-test: for each competitor $k\neq k^\star$, it forms a standardized gap
$z=(\hat\mu_k-\hat\mu_{k^\star})/\sqrt{\hat\sigma_k^2+\hat\sigma_{k^\star}^2}$ and compares it against a Bonferroni-adjusted critical value $z_{1-\alpha_{\mathrm{adj}}}$ to test whether $k^\star$ is significantly faster than $k$. If all pairwise tests pass, the algorithm exploits by returning $k^\star$; otherwise it falls back to the same exploration step as $\tool{}^{\mathrm{Bootstrap}}$, executing all kernels on the current context and appending $(\bm q,\bm y)$ to $\mathcal D$.

The practical difference is therefore computational and statistical: $\tool{}^{\mathrm{Bootstrap}}$ is nonparametric and derives its decision from an empirical distribution induced by weighted resampling, whereas $\tool{}^{\mathrm{CLT}}$ is parametric and replaces resampling with a Gaussian approximation and a single z-test per competitor (removing the bootstrap loop, but relying on the adequacy of the CLT approximation under the locality-weighted data). In practice, we rely on $\tool{}^{\mathrm{CLT}}$ for decision making, since $\tool{}^{\mathrm{Bootstrap}}$ requires extensive sampling from a discrete distribution, which would significantly increase inference latency. Surprisingly, in our experiments (Section~\ref{sec:expr}), we found that $\tool{}^{CLT}$ often outperformed $\tool{}^{Bootstrap}$ on accuracy (not just inference speed). This could imply that the Gaussian approximation is actually a regularizing term, and that the behavior of kernels in the local neighborhood in feature space is, in fact, approximately Gaussian.

\sparagraph{Integration with MDP} \tool{} integrates naturally with morsel-driven parallelism (MDP)~\cite{morsel}. Each worker thread maintains its own local experience buffer and makes kernel-selection decisions using only this local state, allowing morsels to be processed without any additional synchronization beyond the engine’s existing scheduling -- thus, \tool{} adds no synchronization on the critical path. After a query completes (or when the system is idle), the per-thread experience buffers are then synchronized. This merge occurs off the critical path and should amortize well in practice, as new records are generated primarily during exploration, which is infrequent once decisions stabilize (see Section~\ref{sec:regret_tree}, next).

\subsection{Compiled regret trees}
\label{sec:regret_tree}

During early execution, some inference overhead can be justified for learning (feature extraction, model scoring, exploration logic) because it amortizes against large performance gaps and rapidly changing beliefs. However, once the policy has converged and the relevant feature space is well covered, repeatedly invoking the full inference procedure on every morsel becomes overkill: after convergence, \tool{}'s kernel choice is effectively deterministic.

A tempting simplification is to ``freeze'' the policy by training a standard multi-class classifier that predicts the best kernel with high accuracy on historical data. However, this objective is misaligned with our goals. Accuracy treats all mistakes equally: confusing two kernels whose latencies differ by 0.1ms is penalized the same as confusing kernels that differ by 10ms. For kernel selection, what matters is not whether the predicted label is exactly correct, but how much extra time is incurred when it is wrong. The frozen policy should therefore prioritize correctness in regions where mistakes are expensive, and tolerate ambiguity where kernels are nearly tied.

We address this with \emph{regret trees}, a cost-aware decision tree~\cite{cart} trained to minimize empirical \emph{regret} rather than misclassification rate. This approach is enabled by counterfactual measurements: for a subset of morsels: for each training morsel with feature vector $\bm x_i$ and observed latencies ${y_{i,k}}_{k\in\mathcal K}$, define the per-arm regret \[r_{i,k} =y_{i,k} - \min_{k'\in\mathcal K} y_{i,k'}\] which quantifies the time penalty of choosing kernel $k$ instead of the best observed alternative. A regret tree learns a mapping $T:\bm x \mapsto k$ by minimizing the total regret $\sum_i r_{i,T(\bm x_i)}$. Concretely, for any node containing a subset $S$ of training points, the optimal leaf action is the kernel with minimum cumulative regret on that subset, $k_S=\arg\min_k \sum_{i\in S} r_{i,k}$, and the node’s loss is \[\mathcal L(S)=\min_k \sum_{i\in S} r_{i,k}\] Splits are chosen to maximize regret reduction, i.e., to minimize $\mathcal L(S_\text{left})+\mathcal L(S_\text{right})$, yielding partitions that are driven by \emph{where the regret is large} rather than where the label boundary is sharp.

\begin{figure}
    \centering
    \includegraphics[width=\linewidth]{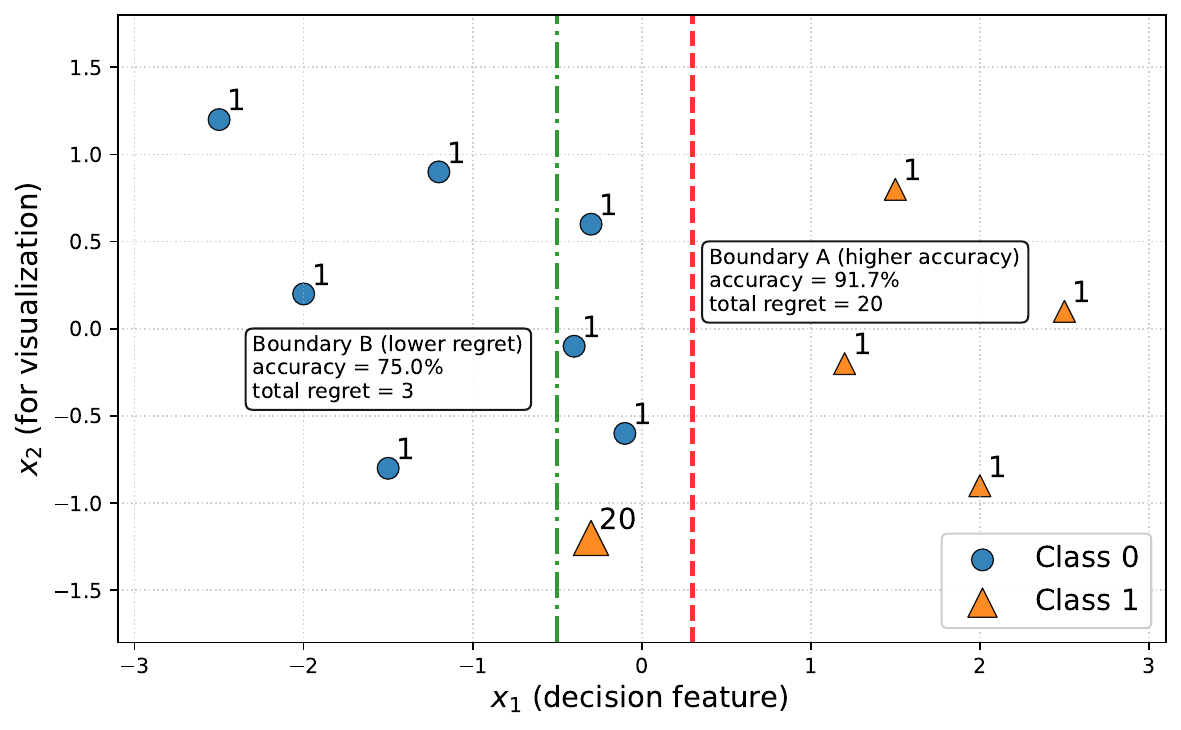}
    \caption{Minimizing regret vs maximizing accuracy. Each point is annotated with the penalty of misclassification.}
    \label{fig:regret}
\end{figure}

\sparagraph{Example} \autoref{fig:regret} illustrates the difference between standard decision tree boundary selection and regret tree boundary selection. The number beside each point denotes the regret incurred when that point is misclassified. Although boundary A achieves higher classification accuracy, it misclassifies a point with a large performance gap between different kernels, leading to worse end-to-end performance. In contrast, boundary B has lower accuracy but incurs substantially smaller regret, resulting in better overall performance. This “compiled” policy has negligible runtime cost (a handful of comparisons and a direct jump to the selected kernel), while preserving the end-to-end objective.

\sparagraph{When should a regret tree be built?} Ideally, a regret tree should be built once the model has converged, and has nothing left to learn. Unfortunately, detecting model convergence is not trivial, although there are several practical techniques available today~\cite{rl_conv1, rl_conv2, rl_conv3, rl_conv4}. Therefore, we leave finding the optimal convergence detection procedure to future work. We show experimentally in Section~\ref{sec:expr} that building a regret tree after the first dozen queries have been processed is often sufficient. Another potential solution is to monitor inference time. Developers can set an upper bound on inference duration, then construct a regret tree when the inference time approaches this limit or if no new points are identified as ``explore'' for an extended period.

%% file: experiments.tex
\section{Experiments}
\label{sec:expr}

To comprehensively understand the ability of \tool{} to improve an existing execution engine, we conducted a series of experiments to answer the following research questions:
\begin{itemize}
    \item Can \tool{} improve the end-to-end query time (i.e., taking learning overhead into account)?
    \item What does the overhead of \tool{} look like?
    \item How well does the CLT approximation work?
    \item How sensitive is \tool{} to the hyperparameter values?
    \item How sample efficient is \tool{}?
    \item Can the regret tree optimization reduce inference costs without compromising performance?
\end{itemize}

Next, we describe our experimental setup and prototype implementation of \tool{}.

\begin{table*}
  \renewcommand\arraystretch{1.12}
  \caption{Tasks evaluated with \tool{}}
  \begin{tabularx}{\linewidth}{c c X}
    \toprule
    Tasks & Options & Comments\\
    \midrule
    \multirowcell{4}{Filter Iterator \\Selection}
        & Index Iterator & Enumerate true-bit positions and gather corresponding elements into new contiguous buffers. oracle for sparse predicates where per-index gathers beat range copies.\\
        & Slice Iterator & Collapse the boolean mask into contiguous true runs and, for each, bulk-copy those spans from the source into the output. Best for dense masks with long true runs or when most rows are kept.\\
    \midrule
    \multirowcell{4}{Sorting Algorithm \\Selection}
        & Quicksort & Efficiency depends strongly on how the input data are distributed. Performs well when the data allow balanced partitions (e.g., random or well-shuffled inputs). \\
        & Heapsort & Runtime is largely insensitive to the input data distribution, and thus does not benefit from favorable input distributions. \\
    \midrule
    \multirowcell{4}{Nested Predicate\\Evaluation (NPE)}
        & Parallel & Run both predicates to get two bitmaps, then intersect two bitmaps to get the final result. It performs well when many rows pass the inner predicate.\\
        & Sequential & Compute the inner predicate, then check the other predicate only for valid rows to avoid scanning unselected data—good when the base selection is sparse or outer predicate is costly to evaluate.\\
    \bottomrule
  \end{tabularx}
  \label{tab:kernels}
\end{table*}

\sparagraph{Database Integration \& Tasks} We build on arrow-rs~\cite{noauthor_apache_nodate}, a widely used columnar data processing library, to implement \tool{}. We integrate \tool{} into three representative tasks (\autoref{tab:kernels}): filter iterator selection (slice-based iteration vs. point-based iteration), sorting algorithm selection (quicksort vs. heapsort), and nested predicate evaluation (partial vs. independent evaluation).

\sparagraph{Baselines} We compare \tool{} with several baselines. First, we include the best single algorithm (``SingleBest'') for each task (i.e., the best kernel to pick if you could only pick one). Second, we include a clairvoyant, zero-latency kernel selector (``Oracle''), which was computed by enumerating all options beforehand (not an option in practice). Note that, because of variance, ``Oracle'' can be slightly outperformed if another technique experiences lucky variance. Third, we include human-authored baselines (``Heuristic''), which we tuned for our hardware platform. Fourth, we include a non-contextual multi-armed bandit (MAB) algorithm (UCB), identical to the implementation in Excalibur~\cite{excalibur} (see Section~\ref{sec:rw}). We also experimented with state-of-the-art RL algorithms (e.g.,\cite{ppo}); however, because their inference overhead exceeds kernel execution time, we were unable to achieve a reasonable result.

\sparagraph{Hyperparameter Tuning} We tuned the hyperparameters of \tool{} and all baselines using Bayesian optimization via the Tree-structured Parzen Estimator~\cite{watanabe_tree-structured_2025} (TPE), using the Optuna~\cite{akiba_optuna_2019} implementation with the objective of minimizing overall regret. This optimization is performed offline and introduces no runtime overhead.

\sparagraph{Platform \& Experimental Methodology} All experiments are conducted on an AMD Ryzen 7 Pro 8700GE CPU and 128 GB of RAM. Unless otherwise noted, each experiment was repeated five times. For deterministic algorithms (Heuristic and SingleBest), we plot the median latency. For stochastic methods, we display the average latency with a bootstrapped 95\% confidence interval.

\sparagraph{Workload and Datasets}
We evaluate \tool{} on three datasets:

\begin{itemize}
\item \textbf{IMDb:} Introduced in~\cite{howgood}, this dataset contains metadata about films. Its non-uniform data distributions make it well-suited for evaluating \tool{} under realistic, skewed workloads.
\item \textbf{Stack:} A subset of the StackOverflow benchmark from Bao~\cite{bao}, containing questions and answers from StackExchange.
\item \textbf{DSB}: The DSB dataset~\cite{ding_dsb_2021}, which is a modification of TPC-DS~\cite{tpcds} with complex data distributions to benchmark both learned and traditional database systems.
\end{itemize}

Since the original benchmark queries associated with each dataset do not all test our three tasks (Table~\ref{tab:kernels}), we used the technique from SQLStorm~\cite{sqlstorm} to generate representative queries with large language models (LLMs). For each dataset, we generated queries until we reached a set of 500 queries that each triggered at least one of the considered tasks. Our datasets and queries are available in the GitHub repository linked on page 1.

\subsection{Performance Improvements}
Here, we measure \tool{}'s ability to improve the end-to-end performance of our workloads.

\begin{figure*}
    \centering

    \begin{subfigure}{\textwidth}
        \centering
        \begin{tabular}{@{}c c@{}}
            \raisebox{0.5\height}{\rotatebox{90}{\Large\bfseries IMDb}} &
            \includegraphics[width=0.92\textwidth]{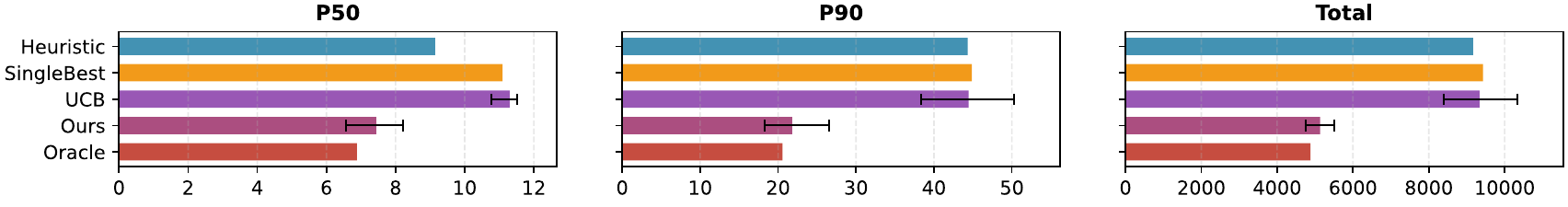}
        \end{tabular}
    \end{subfigure}

    \begin{subfigure}{\textwidth}
        \centering
        \begin{tabular}{@{}c c@{}}
            \raisebox{0.6\height}{\rotatebox{90}{\Large\bfseries Stack}} &
            \includegraphics[width=0.92\textwidth]{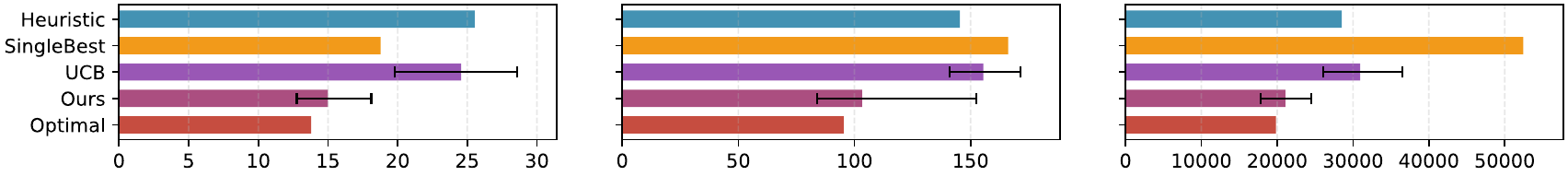}
        \end{tabular}
    \end{subfigure}

    \begin{subfigure}{\textwidth}
        \centering
        \begin{tabular}{@{}c c@{}}
            \raisebox{1.3\height}{\rotatebox{90}{\Large\bfseries DSB}} &
            \includegraphics[width=0.92\textwidth]{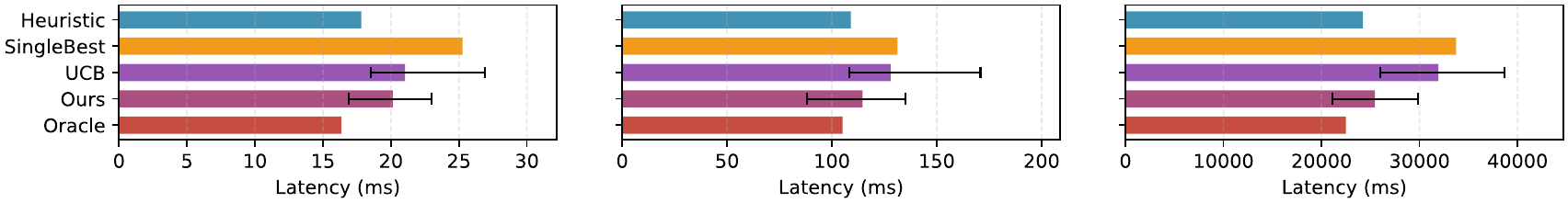}
        \end{tabular}
    \end{subfigure}
    \caption{Latency comparison across datasets}
        \label{overall-latency}
\end{figure*}

\begin{figure*}
\vspace*{-0.5cm}
     \begin{subfigure}{\linewidth}
        \centering
        \includegraphics[width=0.8\linewidth]{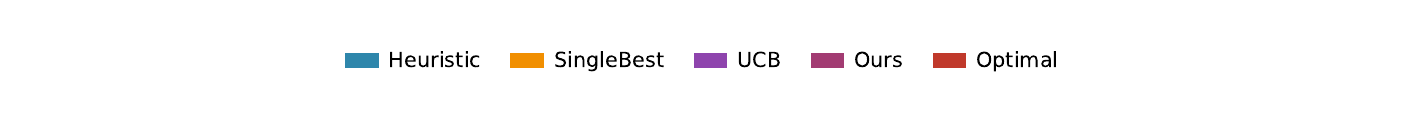}
        \caption*{}
        \vspace{-10mm}
    \end{subfigure}
    \begin{subfigure}{0.32\linewidth}
        \centering
        \includegraphics[width=\linewidth]{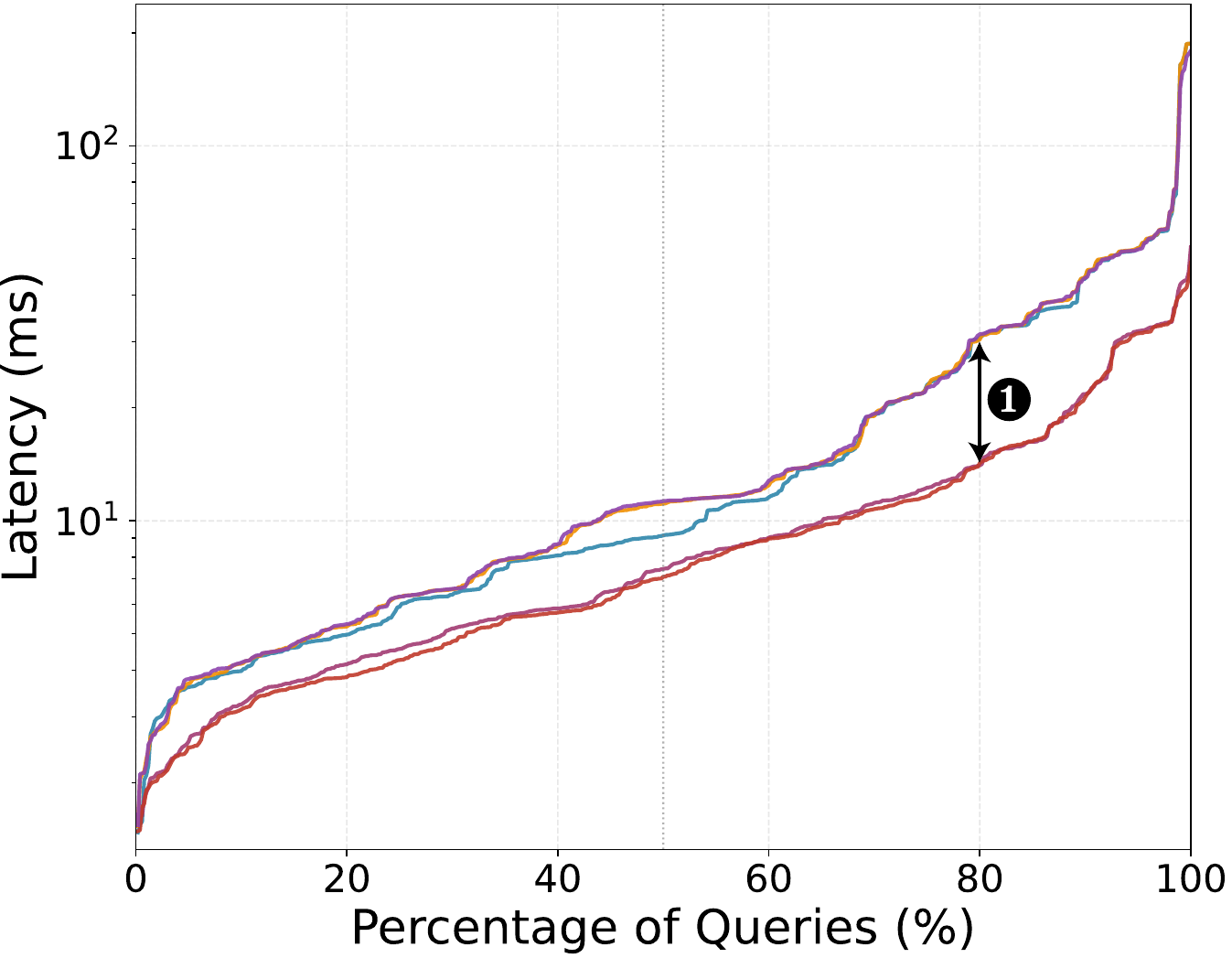}
        \caption{IMDb}
    \end{subfigure}
    \begin{subfigure}{0.32\linewidth}
        \centering
        \includegraphics[width=\linewidth]{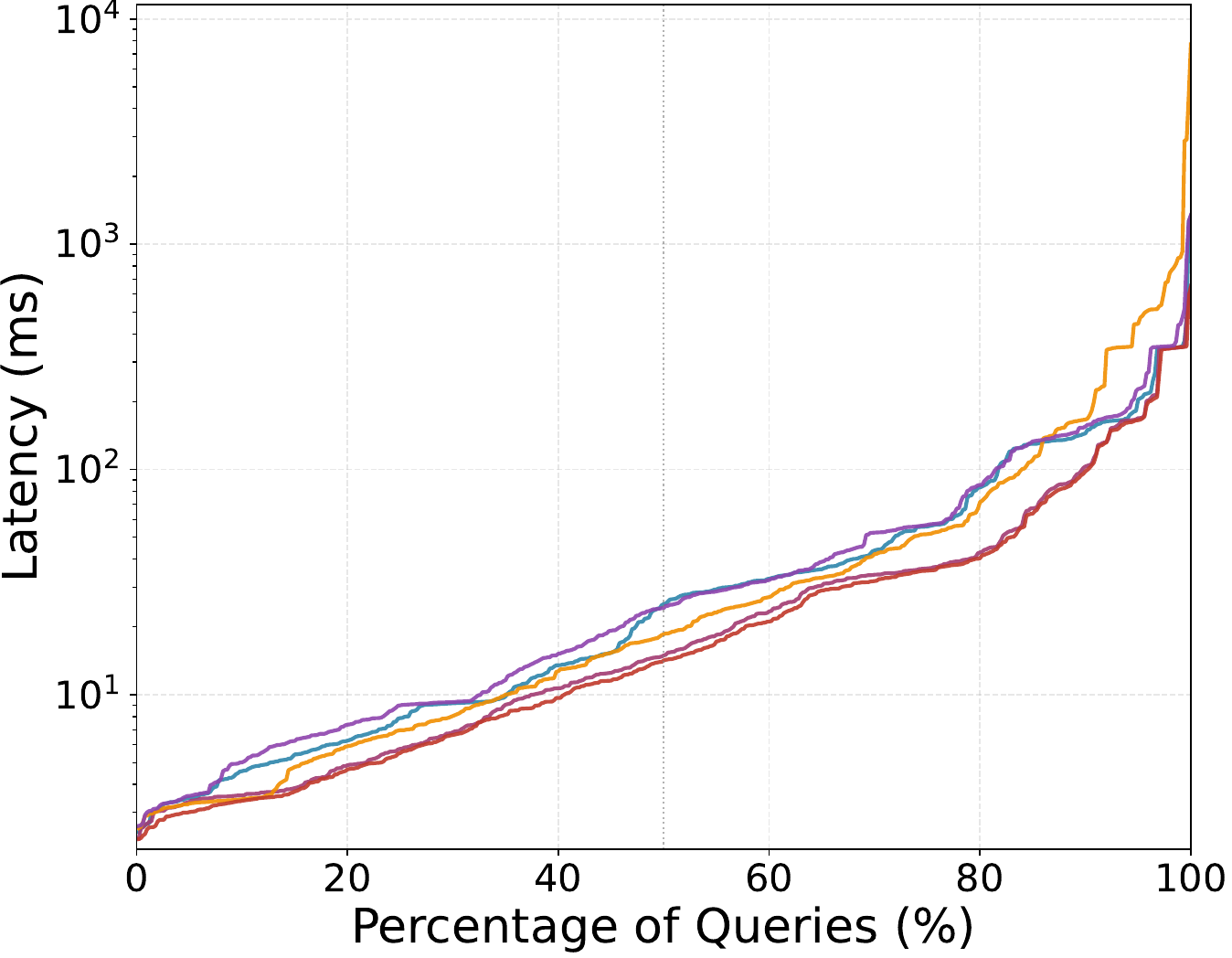}
        \caption{Stack}
    \end{subfigure}
    \begin{subfigure}{0.32\linewidth}
        \centering
        \includegraphics[width=\linewidth]{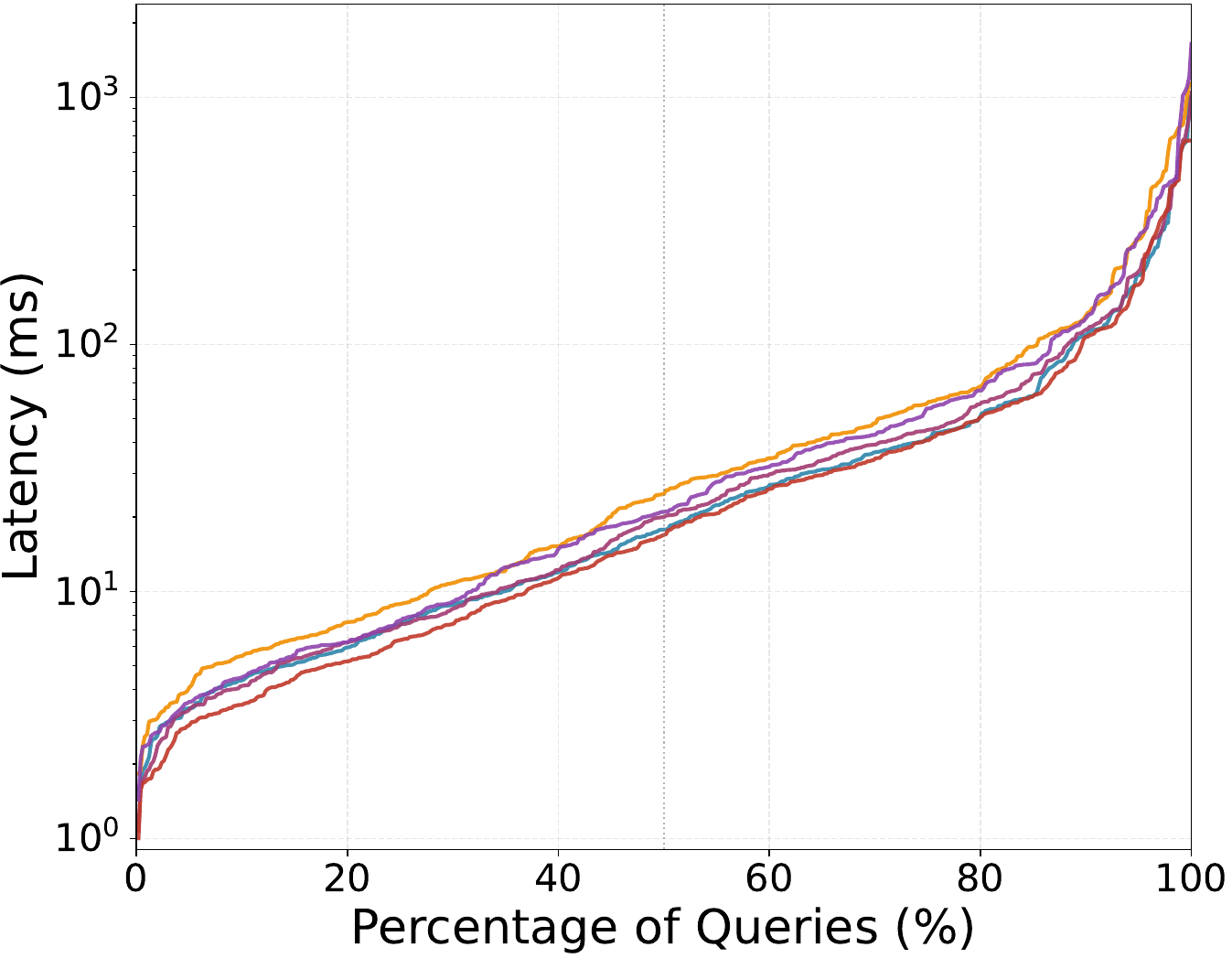}
        \caption{DSB}
    \end{subfigure}
    \caption{Cumulative Distribution Function (CDF) for query latencies. A gap between the curves indicates strictly better performance (at all quantiles) of the lower curve. \circleOne \ shows the performance gain of our method over the heuristic baseline, achieving performance nearly indistinguishable from the optimal oracle at P80. }

    \label{fig:cdf}
\end{figure*}

\sparagraph{Reducing End‑to‑End Latency.} \autoref{overall-latency} shows the median (P50), P90, and total query latencies for our test workloads using different kernel selection algorithms. CDFs are shown in \autoref{fig:cdf}.

On the \textbf{IMDb} dataset, \tool{} yields the largest tail and total improvements and is close to the performance of the oracle overall. Comparing with the heuristic, \tool{} cuts P90 from \qty{44.20}{\milli\second} to \qty{21.77}{\milli\second} (-50.7\%). \tool{} also reduces the worst-case query from 187.29 ms (Heuristic) to 53.59 ms (-71.4\%), while remaining within 2.57\% of oracle in total latency. On the \textbf{Stack} dataset, \tool{} consistently improves median, tail, and total latency and closely tracks the oracle policy. \tool{} reduces P90 from 144.61 ms to 103.18 ms (-28.7\%). Relative to oracle, \tool{} is within 5.27\%/5.09\%/4.35\% on P50/P90/total. On the \textbf{DSB} dataset, \tool{} mainly improves over SingleBest and UCB, and gives comparable results against Heuristic. Compared with UCB, \tool{} reduces total latency by 20.3\%. However, compared to Heuristic, \tool{} slightly increases P50 from 17.81 to 19.99 ms and increases total latency by 4.95\%. Nevertheless, \tool{} is still only 10.3\% above Oracle in total latency.

Overall, we observed significant improvements over SingleBest, providing strong evidence that \emph{dynamically switching kernels can achieve better performance than using a fixed kernel with the overall‑best performance.} Compared with the human‑defined heuristics, \tool{} still shows significant advantages on two datasets and achieves comparable performance on the remaining dataset, \emph{without requiring developers to manually write or maintain a heuristic.}

\begin{table*}
\caption{Average latency (per query) breakdown for different learning algorithms on the IMDb dataset. The parenthesized values represent a 95\% confidence interval.}
\begin{tabular}{rcccc}
\toprule
                   & $\textbf{\tool{}}^\text{bootstrap}$ & $\textbf{\tool{}}^\text{CLT}$ & UCB & Heuristic\\
\midrule
Counterfactual     &  \makecell{\textbf{0.12} (0.09, 0.15)}   &     \makecell{\textbf{0.11} (0.08, 0.14)}    &  \makecell{\textbf{0}}  & \makecell{\textbf{0}}\\
Inference          &  \makecell{\textbf{140.22} (124.34, 157.30)} & \makecell{\textbf{0.24} (0.22, 0.27)}    &  \makecell{\textbf{0.04} (0.03, 0.04)}  &\makecell{\textbf{0.01}}\\
Kernel Execution   &  \makecell{\textbf{9.26} (8.58, 10.02)}  &   \makecell{\textbf{8.12} (7.55, 8.76)}    &  \makecell{\textbf{16.87} (15.04, 18.92)}   &\makecell{\textbf{16.56}}\\
End-to-End &  \makecell{\textbf{149.68} (133.19, 167.37)} & \makecell{\textbf{8.55} (7.95, 9.21)}    &   \makecell{\textbf{16.98} (15.15, 19.04)} & \makecell{\textbf{16.57}} \\
\bottomrule
\end{tabular}
\label{tab:approx}
\end{table*}

\sparagraph{CLT approximation}  In Section~\ref{sec:cake}, we described \tool{}'s bootstrap variant (\bootstrap), along with the optimized and approximate version (\clt). Here, we assess how well the CLT optimization works. \autoref{tab:approx} illustrates the performance, accuracy, and overhead differences between \bootstrap and \clt. While \bootstrap attains low total kernel execution time (\qty{9.26}{\milli\second}), its bootstrap-based inference introduces a prohibitive \qty{140.22}{\milli\second} overhead cost per query (due to the cost of drawing each bootstrap sample). Replacing bootstrap sampling with the CLT approximation reduces the total inference overhead to \qty{0.24}{\milli\second} while slightly improving kernel execution to \qty{8.12}{\milli\second}, resulting in an \qty{8.55}{\milli\second} end-to-end latency. This is about a $2\times$ speedup over UCB (16.98 ms) and the heuristic baseline (16.57 ms). We hypothesize two potential reasons for the better performance of \clt compared to \bootstrap. First, \clt forces the variance of each sample to be Gaussian (whereas \bootstrap would only converge to a Gaussian after many iterations), and thus acts as a form of regularization, reducing the variance of the model. Second, sampling from a multinomial distribution requires constructing and accessing additional data structures, which can interfere with prefetching mechanisms and degrade performance.

\begin{table*}[htbp!]
\caption{Average latency (per query) breakdown with and without compilation on the IMDb dataset. The parenthesized values represent a 95\% confidence interval.}
\resizebox{\textwidth}{!}{%
\begin{tabular}{lcccccccc}
\toprule
        & \multicolumn{4}{c}{\textbf{Without Compilation}}                                                                                          & \multicolumn{4}{c}{\textbf{With Compilation (Regret Tree)}}                                                                                               \\ \cmidrule(lr){2-5} \cmidrule(lr){6-9}
        & \multicolumn{1}{c}{Counterfactual} & \multicolumn{1}{c}{Inference} & \multicolumn{1}{c}{Kernel} & \multicolumn{1}{c}{End-to-End} & \multicolumn{1}{c}{Counterfactual} & \multicolumn{1}{c}{Inference} & \multicolumn{1}{c}{Kernel}   & \multicolumn{1}{c}{End-to-End} \\ \midrule
    Epoch 1 & \textbf{0.11} (0.08, 0.14)  & \textbf{0.22} (0.21, 0.25)  & \textbf{8.05} (7.48, 8.69) & \textbf{8.37} (7.86, 9.11) & \textbf{0.11} (0.08, 0.14)   & \textbf{0.22} (0.20, 0.24)    & \textbf{8.07} (7.49, 8.71) & \textbf{8.4 }(7.87, 9.14)   \\
    Epoch 2 & \textbf{0.01} (0.00, 0.01) & \textbf{0.28} (0.26, 0.31)  & \textbf{8.11} (7.53, 8.76) & \textbf{8.42} (7.86, 9.15) & \cellcolor[HTML]{90EE90}- & \cellcolor[HTML]{90EE90}\textbf{0.01} (0.01, 0.04) & \cellcolor[HTML]{90EE90}\textbf{7.94} (7.37, 8.57)& \cellcolor[HTML]{90EE90}\textbf{7.95} (7.45, 8.68) \\
    Epoch 3 & \textbf{0.00} (0.00, 0.00) & \textbf{0.29} (0.27, 0.32)  & \textbf{8.11} (7.53, 8.76) & \textbf{8.53} (7.87, 9.15) & \cellcolor[HTML]{90EE90}-& \cellcolor[HTML]{90EE90}\textbf{0.01} (0.01, 0.01) & \cellcolor[HTML]{90EE90}\textbf{7.94} (7.37, 8.57)& \cellcolor[HTML]{90EE90}\textbf{7.95} (7.44, 8.66) \\
    Epoch 4 & \textbf{0.00} (0.00, 0.00) & \textbf{0.29} (0.27, 0.32)  & \textbf{8.12} (7.53, 8.76) & \textbf{8.53} (7.87, 9.16) & \cellcolor[HTML]{90EE90}-& \cellcolor[HTML]{90EE90}\textbf{0.01} (0.01, 0.01) & \cellcolor[HTML]{90EE90}\textbf{7.94} (7.37, 8.58)& \cellcolor[HTML]{90EE90}\textbf{7.95} (7.44, 8.67) \\
    Epoch 5 & \textbf{0.00} (0.00, 0.00) & \textbf{0.29} (0.27, 0.32)  & \textbf{8.11} (7.52, 8.75) & \textbf{8.53} (7.86, 9.15) & \cellcolor[HTML]{90EE90}-& \cellcolor[HTML]{90EE90}\textbf{0.01} (0.01, 0.01) & \cellcolor[HTML]{90EE90}\textbf{7.94} (7.37, 8.58)& \cellcolor[HTML]{90EE90}\textbf{7.95} (7.45, 8.67) \\ \bottomrule
\end{tabular}
}
\label{tab:compilation}
\end{table*}

\sparagraph{Regret tree} Since \tool{} can get many samples from a single query, \tool{}'s models can converge quickly. Once \tool{}'s models are converged, we can compile them into efficient regret trees with even lower inference time. To test this capability, we ran the IMDb workload in a loop. We refer to each loop of the workload as an epoch. \autoref{tab:compilation} illustrates the impact of the regret tree compilation strategy on latency. Without compilation, inference latency increases until the history buffer reaches its maximum capacity during Epoch 2, stabilizing at 0.28 ms per query. In contrast, using the regret tree reduces inference time to \qty{0.01}{\milli\second} per query, accompanied by a slight decrease in kernel execution time (likely due to \tool{}'s regret tree using fewer registers/cache). These results demonstrate that regret trees can significantly mitigate inference overhead without compromising, and potentially enhancing, overall performance.

\subsection{Quantified Overheads}
\begin{table}[htbp!]
\caption{Average latency (per query) breakdown. For each dataset, \tool{} spends around 95\% of the time executing kernels. Counterfactual execution and model-related overhead is around 5\% (without regret tree compilation).}
\resizebox{0.98\linewidth}{!}{%
\begin{tabular}{lccc}\toprule
                & \textbf{Model Related}                     & \textbf{Kernel Execution}                     & \textbf{Counterfactual}                     \\ \midrule
\textbf{IMDb}            & \makecell{3.76\% (0.32 ms/q)} & \makecell{95.00\% (8.12 ms/q)}  & \makecell{1.24\% (0.11 ms/q)} \\
\textbf{Stack}           & \makecell{4.93\% (1.05 ms/q)} & \makecell{94.23\% (20.09 ms/q)} & \makecell{0.84\% (0.18 ms/q)} \\
\textbf{DSB}             & \makecell{5.45\% (1.76 ms/q)} & \makecell{94.12\% (30.48 ms/q)} & \makecell{0.42\% (0.14 ms/q)} \\ \bottomrule
\end{tabular}
}
\label{latency_breakdown}
\end{table}

The overhead incurred by \tool{} during the online‑learning stage is attributable to three components: feature extraction, conditional sampling (inference), and the generation of counterfactuals. After \tool{} transitions to the predict‑only phase (the regret tree), the only residual overhead stems from feature extraction.

\autoref{latency_breakdown} presents a breakdown of the end‑to‑end latency incurred by \tool{} during the online learning stage. Across all three datasets, the total overhead accounts for only 5\% of the end‑to‑end latency. In contrast to the substantial performance gains achieved by the optimized kernels, this trade‑off yields a clearly beneficial improvement (see Figure~\ref{overall-latency}). Moreover, the relatively small counterfactual overhead demonstrates that generating only a small number of counterfactuals is sufficient to improve model performance, thereby validating the underlying assumption of our algorithm.

\begin{figure}[htbp!]
    \centering
        \hspace*{-0.6cm}
    \includegraphics[width=0.8\linewidth]{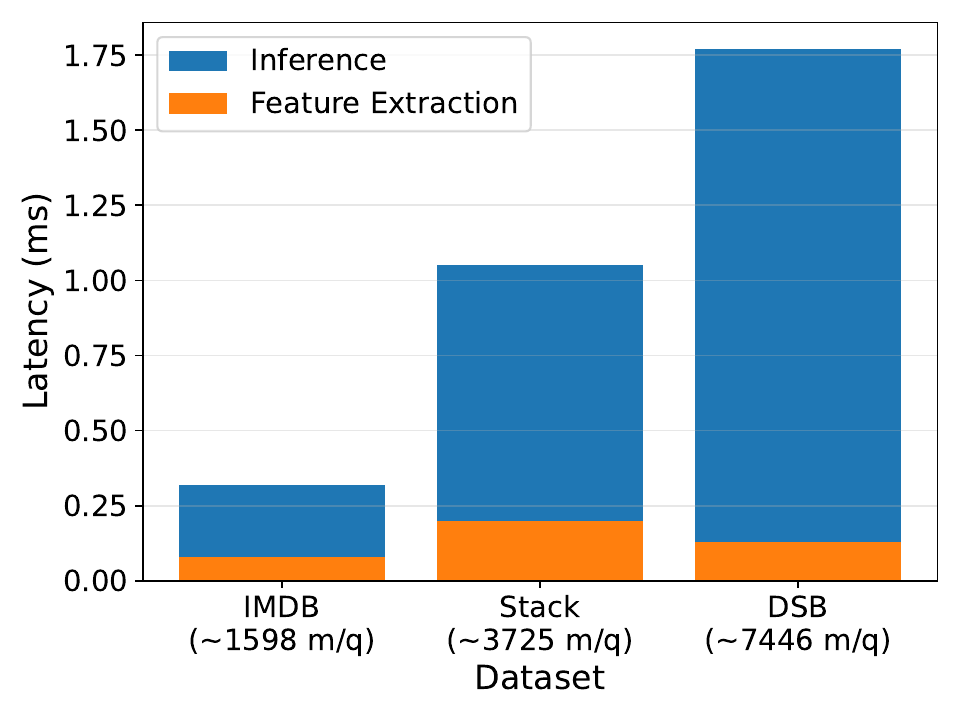}
    \caption{Per-query model related overhead, broken down by inference and feature extraction time. The number of morsels per query (m/q) is listed under each workload name. Datasets with more morsels per query have higher overhead, since \tool{} is invoked more times. Model inference represents the bulk of the model-related overhead.}
    \label{fig:inference_overhead}
\end{figure}

\autoref{fig:inference_overhead} presents a breakdown of the average model‑related overhead for each dataset. Two observations can be drawn from the results. First, the overall overhead of \tool{} is negligible on a per-query basis. For the IMDb data set, the model is invoked to process an average of 1,598 morsels per query, yet the incurred overhead amounts to only 0.28 ms per query. This corresponds to an average model‑related latency of ~175 ns per morsel, which has a minimal impact on the end‑to‑end latency. Secondly, the inference time per query grows almost linearly with the number of morsels per query, rather than quadratically. A naive worst‑case analysis might suggest a quadratic time complexity: inference time scales linearly with the amount of data retained in the history buffer, and, under the assumption of a constant sampling rate $p$, one would obtain a complexity of $O(p \cdot n \cdot n)=O(n^{2})$ where n denotes the number of morsels to process. However, our empirical results show that the number of counterfactuals actually generated is very small, as exactly the pattern reported in Figure~\ref{latency_breakdown}. Consequently, the observed inference overhead behaves much closer to $O(n)$ in practice.

\begin{figure}[htbp!]
    \centering
    \hspace*{-0.5cm}
    \includegraphics[width=0.8\linewidth]{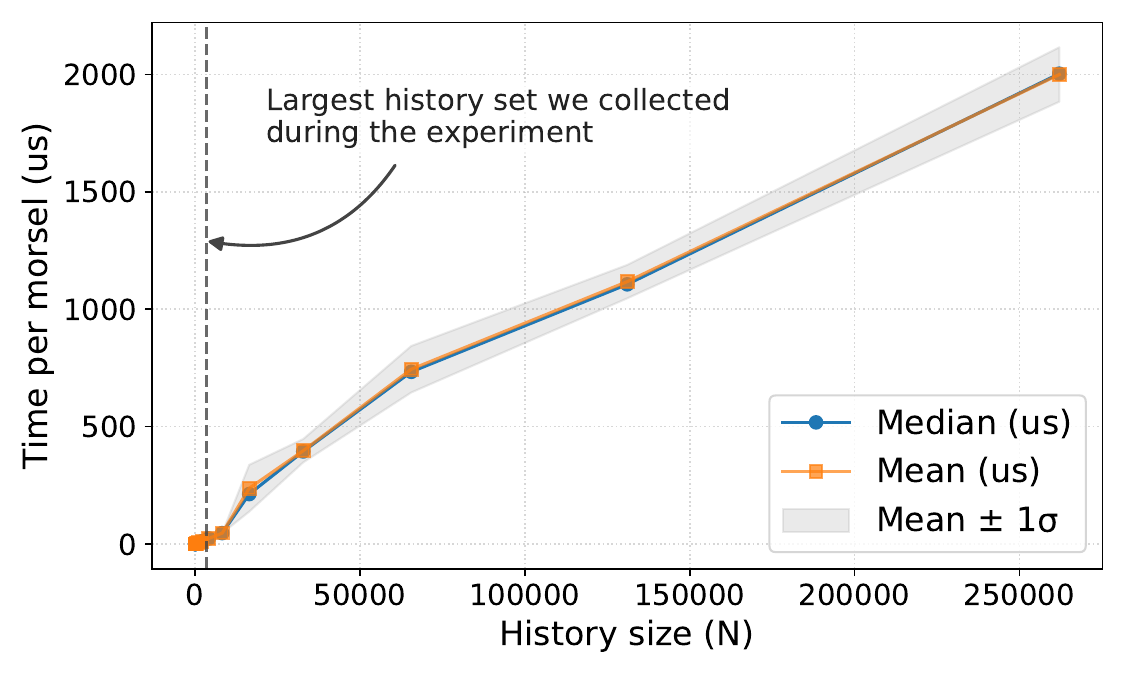}
    \caption{Inference overhead growth with increasing history size. If the history buffer grows unbounded, inference latency will increase. The vertical line represents the largest history buffer observed in our experiments.}
    \label{fig:inference_growth}
\end{figure}

In \autoref{fig:inference_growth}, we analyze how \tool{}'s inference time scales with the size of the history buffer. As the amount of data stored in the history buffer increases, the time required for a single conditional‑sampling operation grows linearly. Nevertheless, for a reasonably sized history buffer, the inference latency remains low.
Thus, \tool{} is particularly well-suited for time-constrained scenarios, especially in contrast to the typical millisecond-level inference latency of other popular RL algorithms~\cite{thodoroff_benchmarking_2022}.
Maintaining a bounded history buffer is common in other applications of RL to database systems as well~\cite{bao}.

\begin{figure}
    \centering
    \hspace*{-1cm}
    \includegraphics[width=0.9\linewidth]{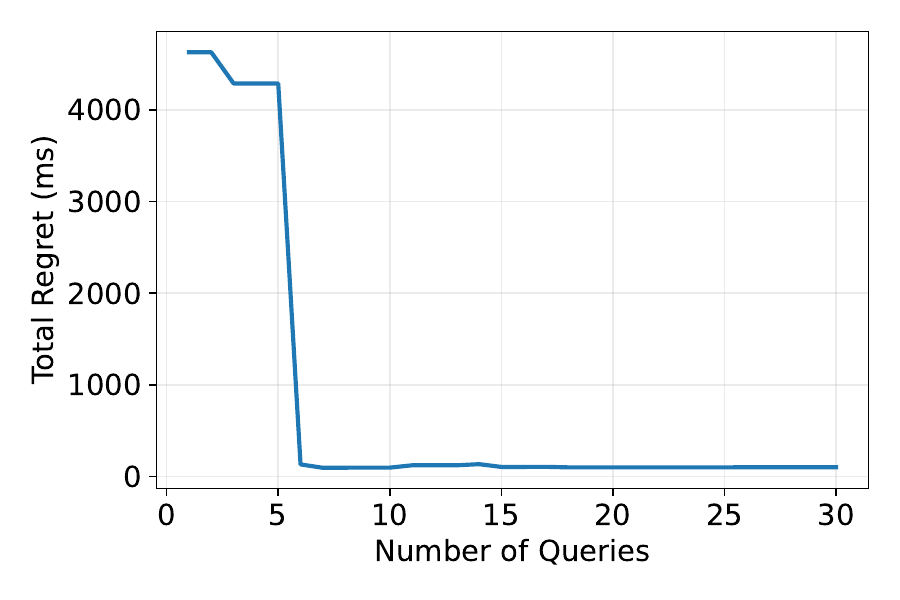}
    \caption{Regret incurred over the complete IMDb dataset when we constrain the learning within first $n$ queries}
    \label{fig:sample_eff}
\end{figure}

\subsection{Sample Efficiency}
To assess sample efficiency, we measure regret while limiting \tool{}'s learning to an initial budget of $n$ queries, then immediately construct the regret tree, reporting the resulting regret incurred over the complete dataset. The results are plotted in \autoref{fig:sample_eff}. The total regret across the entire dataset drops sharply between the fifth and sixth queries, after which it remains largely stable. When \tool{} learns from only six queries, it processes approximately 8,000-10,000 morsels during this window. Thus, \tool{} does not suffer from the same sample efficiency issues that are endemic in other RL-based database optimization techniques (e.g.,~\cite{neo}).

\begin{figure}
    \centering
    \includegraphics[width=\linewidth]{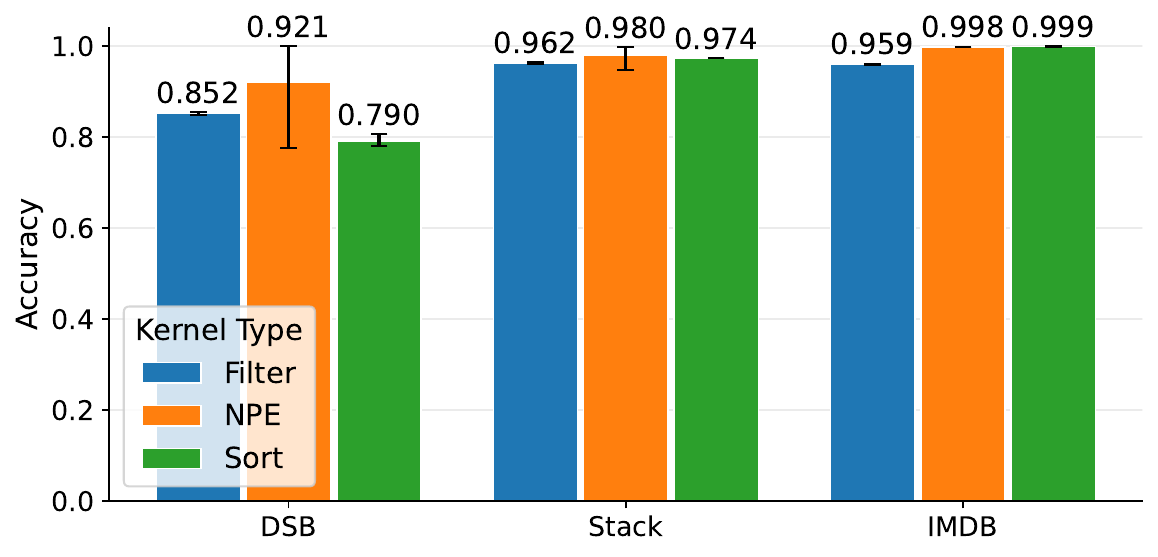}
    \caption{Accuracy of \tool{} for different kernels}
    \label{fig:kernel_acc}
\end{figure}

\subsection{Kernel Analysis}

To understand where \tool{}’s end-to-end gains originate, we examine decision quality at the granularity of individual kernel selections for the three tasks in \autoref{tab:kernels}. Figure~\ref{fig:kernel_acc} summarizes kernel-selection accuracy (compared to the Oracle, which has 100\% accuracy) across datasets during the initial run. On IMDb, \tool{} reaches 95.9\% accuracy for filter selection, and is nearly perfect for NPE and sort selection (99.8\% and 99.9\%). On Stack, accuracies are similarly high (96.2\%/98.0\%/97.4\% for filter/NPE/sort). DSB is more challenging: while NPE remains accurate (92.1\%), filter and sort selection accuracy drop to 85.2\% and 79.0\%, respectively, aligning with the smaller end-to-end gains observed on DSB relative to the hand-coded heuristic. This may be attributable to the fact that DSB is designed to be adversarial to learned models~\cite{ding_dsb_2021}.

\begin{table*}
\centering
\label{tab:combined_confusion_matrix}
\caption{Confusion matrices for each task on the IMDb dataset.}
\resizebox{0.9\textwidth}{!}{%

\begin{subtable}[b]{0.32\textwidth}
    \centering
    \caption{Query Plan}
    \begin{tabular}{@{}l|ccc@{}}
    \toprule
    \diagbox[width=6.2em]{\textbf{Actual}}{\textbf{Pred}} & \textbf{Para.} & \textbf{Seq.} & \textbf{Total} \\ \midrule
    Para. & 30,187 & 85 & 30,272 \\
    Seq.  & 446 & 89,835 & 90,281 \\ \midrule
    \textbf{Total} & 30,633 & 89,920 & 120,553 \\ \bottomrule
    \end{tabular}
\end{subtable}\hspace{0.6em}
\begin{subtable}[b]{0.32\textwidth}
    \centering
    \caption{Filter}
    \begin{tabular}{@{}l|ccc@{}}
    \toprule
    \diagbox[width=6.2em]{\textbf{Actual}}{\textbf{Pred}} & \textbf{Slice} & \textbf{Index} & \textbf{Total} \\ \midrule
    Slice & 92,255 & 57,322 & 149,577 \\
    Index & 9,957 & 579,457 & 589,414 \\ \midrule
    \textbf{Total} & 102,212 & 636,779 & 738,991 \\ \bottomrule
    \end{tabular}

\end{subtable}\hspace{0.6em}
\begin{subtable}[b]{0.32\textwidth}
    \centering
    \caption{Sort}
    \begin{tabular}{@{}l|ccc@{}}
    \toprule
    \diagbox[width=6.2em]{\textbf{Actual}}{\textbf{Pred}} & \textbf{Quick} & \textbf{Heap} & \textbf{Total} \\ \midrule
    Quick & 39,350 & 1,974 & 41,324 \\
    Heap  & 1,716 & 858 & 2,574 \\ \midrule
    \textbf{Total} & 41,066 & 2,832 & 43,898 \\ \bottomrule
    \end{tabular}
\end{subtable}
}
\label{tab:confusion}
\end{table*}

\autoref{tab:confusion} further characterizes the error distribution. For NPE, mispredictions are rare: only 85 Parallel morsels are classified as Sequential and 446 Sequential morsels are classified as Parallel out of 120,553 decisions. For filter selection, errors are asymmetric: Slice is misclassified as Index 57,322 times, whereas the reverse error occurs 9,957 times; this coincides with strong class imbalance, as Index is optimal for 589,414 instances compared to 149,577 for Slice. For sorting, the Heap case is infrequent (2,574 of 43,898 instances), and most errors correspond to confusion between these rare Heap instances and the dominant QuickSort cases.

Finally, we note that while accuracy provides a convenient, task-agnostic summary, it is not \tool{}'s optimization goal: \tool{} targets correctness in parts of the feature space where the regret is high, rather than treating all misclassifications equally.

\begin{table*}
\caption{Average query latency under different levels of hyperparameter optimization. Parenthesized values represent 95\% confidence intervals. Tuning a single set of hyperparameters for all datasets (``BO for all'') has similar performance as tuning individual hyperparameters for each dataset (``BO for each'').}
\vspace*{3pt}
\begin{tabular}{lccc}
\toprule
            & IMDb & Stack & DSB \\ \midrule
Heuristic   &  16.64 ms &36.22 ms & 30.31 ms \\
SingleBest  & 17.15 ms & 84.36 ms& 49.22 ms\\
BO for all &   8.47 ms (7.86, 9.13)   &    21.51 ms (17.97, 25.95)   & 29.85 ms (24.18, 36.27)  \\
BO for each &  8.56 ms (7.55, 9.22)   &    21.32 ms (17.80, 25.76)   & 32.38 ms (26.38, 39.18) \\ \bottomrule
\end{tabular}
\label{tab:bo}
\end{table*}

\begin{figure*}
    \centering
    \begin{subfigure}{0.32\linewidth}
        \centering
        \includegraphics[width=\linewidth]{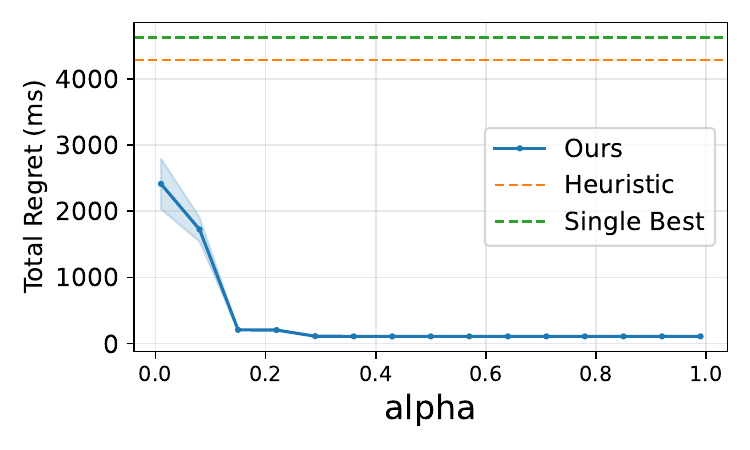}
    \end{subfigure}
    \begin{subfigure}{0.32\linewidth}
        \centering
        \includegraphics[width=\linewidth]{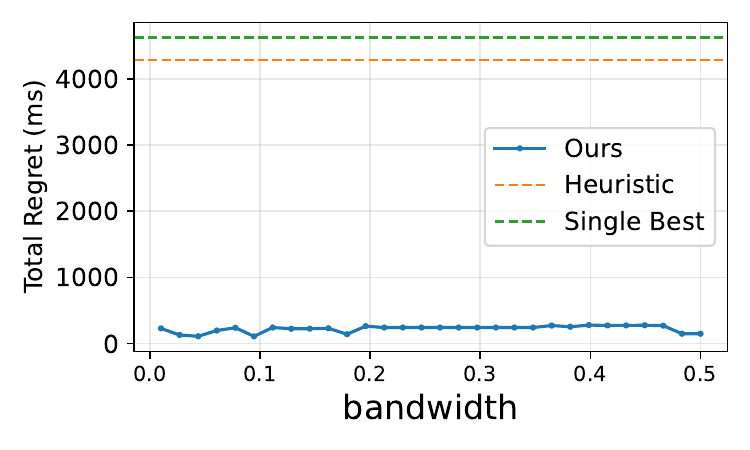}
    \end{subfigure}
    \begin{subfigure}{0.32\linewidth}
        \centering
        \includegraphics[width=\linewidth]{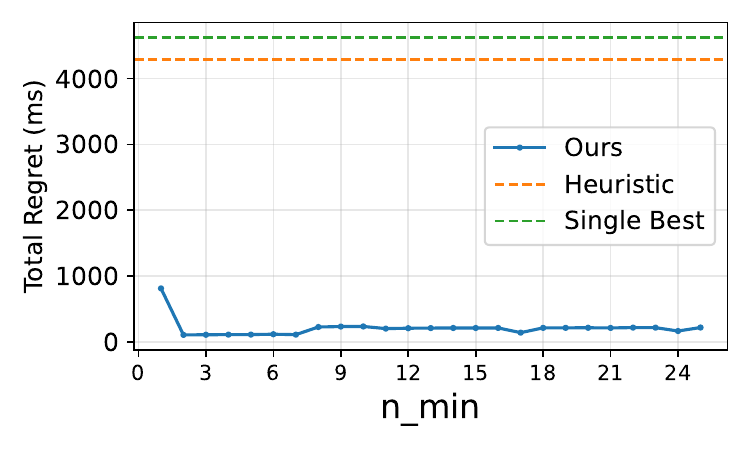}
    \end{subfigure}
    \caption{Hyperparameter sensitivity}
    \label{tab:sweep}
\end{figure*}

\subsection{Hyperparameters}
We empirically evaluated the sensitivity of \tool{} to various hyperparameters. \autoref{tab:bo} presents the average query latency obtained by optimizing hyperparameters under different objectives. Besides the two deterministic baselines (Heuristic and SingleBest), we report the performance of \tool{} under two hyperparameter optimization schemes: (1) \emph{BO for each}, which tunes hyperparameters independently per dataset, and (2) \emph{BO for all}, which performs a single joint optimization across IMDb, Stack, and DSB. We observed that optimizing hyperparameters jointly across all datasets yields performance comparable to that on individual datasets, indicating that tuning the hyperparameters of \tool{} per dataset provides little benefit (i.e., the same hyperparameters can be used across datasets).

We next evaluate sensitivity by sweeping individual hyperparameters around their tuned values while holding others fixed. \autoref{tab:sweep} reports total regret under representative sweeps for three key hyperparameters (alpha, bandwidth, and $n_{\min}$). Across these local ranges, \tool{}’s regret curve remains largely flat, with no sharp degradation, indicating that \tool{} does not rely on narrowly tuned hyperparameters to obtain good decisions. In contrast, both deterministic baselines incur consistently high regret because they do not adapt their kernel choices to the dataset at hand.

Overall, these results suggest that \tool{}’s gains are stable under reasonable hyperparameter variation, and that a single jointly optimized configuration transfers well across workloads.

%% file: relatedwork.tex
\section{Related Work}
\label{sec:rw}

While \tool{} focuses on adapting to different datasets, there is also work seeking to make execution engines adaptive to changes in hardware. To the best of our knowledge, the most similar work is Excalibur~\cite{excalibur} (which built upon micro-adaptivity in Vectorwise~\cite{microadapt}), in which the authors search a wide range of compilation options for kernels written in VOILA~\cite{voila} using two alternative strategies: one that generates candidate variants via heuristics and another that first performs monte carlo tree search to produce variants, and then selecting among the resulting variants using a multi-armed bandit. We compare against Excalibur's multi-armed bandit algorithm in Section~\ref{sec:expr}, but the context-dependent nature of \tool{}'s problem makes a direct comparison impossible. In HAWK~\cite{hawk_adaptive}, the authors propose generating many variants of machine code for the same input kernel, and then learn which variant performs best on the user's hardware. In~\cite{hw_adapt}, the authors propose an MLIR and compilation-based approach. All of these approaches focus on compiling the same logical kernel implementation into different machine code, whereas \tool{} focuses on choosing between different logical kernel implementations. The key difference is that \tool{}'s decisions require context, whereas hardware-based optimization decisions do not (since the hardware does not change query-to-query).

Adaptive query processing (e.g., ~\cite{adaptive_vs_ml,cuttlefish,adaptive_qp_rl,eddies,databricks_adaptive,skinnerdb}) is also relevant, but tends to focus on fixing structural issues with a query plan (e.g., a bad join order) rather than low-level kernel implementations. Query reoptimization~\cite{reopt,workload_reopt,inc_reopt,rio_reopt} similarly seeks to repair structural plan issues by reoptimizing plans mid-execution. 

The most widespread use of RL for database internals is likely RL for query optimization (e.g., ~\cite{neo,balsa,lero,hero_qo,pilotscope,roq,sefrqo,eraser_lqo,fastgres,concurrent_lqo,leon_qo,qo_rank,lstm_jo,bao}). Since most learned query optimizers focus on join ordering, operator selection, and access path selection, \tool{} is mostly orthogonal to these works. Learned query optimizers tend to make all their decisions before each query begins executing, whereas \tool{} actively learns within each query. \tool{} has two key properties that make it easier to use than most learned query optimizers: (1) \tool{} cannot make exceptionally catastrophic decisions, as the difference between the best kernel and the worst kernel is usually a handful of microseconds (whereas the best join order and the worst join order could represent multiple days of computation~\cite{howgood}), and (2) \tool{} gets training data much more quickly than learned query optimizers (i.e., learned query optimizers get one sample per \emph{query}, whereas \tool{} gets one sample per \emph{morsel}), compensating for sample efficiency issues. We leave investigating combining learned query optimizers and \tool{} to future work. 

There is a large amount of work around building adaptive systems. MLOS~\cite{mlos} provides tools for developers to easily benchmark and search the configuration spaces of their programs, but focuses on compile-time tuning (similar to profile-guided optimization~\cite{pgo}). SmartChoices~\cite{smartchoices}, like \tool{}, also allows developers to easily integrate ML-powered decision-making into their applications, but targets latencies in the 10-100 millisecond range, using deep learning and network-attached inference and history buffers. 

\tool{} follows a general theme of building self-tuning and adaptive systems called ``machine programming''~\cite{pillars}. In the context of data management, similar work has been performed for index structures~\cite{ml_index}, cardinality estimation~\cite{deep_card_est2}, query scheduling~\cite{lsched,buffer_sched} and workload management~\cite{wisedb-vldb,auto-wlm,sql_embed}, entity matching~\cite{deep_entity}, and transaction scheduling~\cite{learned_xaction}. 

%% file: conclusion.tex
\section{Conclusion}
\label{sec:conclusion}

This work presented \tool{}, a technique for building adaptive query execution engines using sub-millisecond reinforcement learning techniques. Unlike prior applications of reinforcement learning to database systems, \tool{} learns \emph{within} a query, potentially collecting thousands of samples from one execution. Experimentally, we demonstrated that \tool{} can learn near-optimal policies quickly, dropping tail latency by as much as 2x in our benchmarks. In the future, we plan to investigate how \tool{} can be applied to other elements of the DBMS, such as adaptive query planning~\cite{eddies,databricks_adaptive}, sideways information passing~\cite{yannakakis_debunk,yannakakis_diamond,yannakakis_ml}, cracking kernels~\cite{cracking}, and data layouts~\cite{tsunami,flood,adaptive_science}.

%% file: references.bib
@inproceedings{akiba_optuna_2019,
	address = {New York, NY, USA},
	series = {{KDD} '19},
	title = {Optuna: {A} {Next}-generation {Hyperparameter} {Optimization} {Framework}},
	isbn = {978-1-4503-6201-6},
	shorttitle = {Optuna},
	url = {https://dl.acm.org/doi/10.1145/3292500.3330701},
	doi = {10.1145/3292500.3330701},
	urldate = {2026-01-30},
	booktitle = {Proceedings of the 25th {ACM} {SIGKDD} {International} {Conference} on {Knowledge} {Discovery} \& {Data} {Mining}},
	publisher = {Association for Computing Machinery},
	author = {Akiba, Takuya and Sano, Shotaro and Yanase, Toshihiko and Ohta, Takeru and Koyama, Masanori},
	month = jul,
	year = {2019},
	pages = {2623--2631},
}

@article{nadaraya_estimating_1964,
	title = {On {Estimating} {Regression}},
	volume = {9},
	url = {https://doi.org/10.1137/1109020},
	doi = {10.1137/1109020},
	number = {1},
	journal = {Theory of Probability \& Its Applications},
	author = {Nadaraya, E. A.},
	year = {1964},
	note = {\_eprint: https://doi.org/10.1137/1109020},
	pages = {141--142},
}

@misc{noauthor_apache_nodate,
	title = {Apache {Arrow}},
	url = {https://arrow.apache.org/},
	language = {en-US},
	urldate = {2026-01-30},
	journal = {Apache Arrow},
}

@inproceedings{li_contextual-bandit_2010,
	title = {A {Contextual}-{Bandit} {Approach} to {Personalized} {News} {Article} {Recommendation}},
	url = {http://arxiv.org/abs/1003.0146},
	doi = {10.1145/1772690.1772758},
	urldate = {2026-01-30},
	booktitle = {Proceedings of the 19th international conference on {World} wide web},
	author = {Li, Lihong and Chu, Wei and Langford, John and Schapire, Robert E.},
	month = apr,
	year = {2010},
	note = {arXiv:1003.0146 [cs]},
	pages = {661--670},
}

@inproceedings{thodoroff_benchmarking_2022,
	title = {Benchmarking {Real}-{Time} {Reinforcement} {Learning}},
	issn = {2640-3498},
	url = {https://proceedings.mlr.press/v181/thodoroff22a.html},
	language = {en},
	urldate = {2026-01-28},
	booktitle = {{NeurIPS} 2021 {Workshop} on {Pre}-registration in {Machine} {Learning}},
	publisher = {PMLR},
	author = {Thodoroff, Pierre and Li, Wenyu and Lawrence, Neil D.},
	month = oct,
	year = {2022},
	pages = {26--41},
}

@article{ding_dsb_2021,
	title = {{DSB}: a decision support benchmark for workload-driven and traditional database systems},
	volume = {14},
	issn = {2150-8097},
	shorttitle = {{DSB}},
	url = {https://dl.acm.org/doi/10.14778/3484224.3484234},
	doi = {10.14778/3484224.3484234},
	language = {en},
	number = {13},
	urldate = {2026-01-26},
	journal = {Proceedings of the VLDB Endowment},
	author = {Ding, Bailu and Chaudhuri, Surajit and Gehrke, Johannes and Narasayya, Vivek},
	month = sep,
	year = {2021},
	pages = {3376--3388},
}

@misc{watanabe_tree-structured_2025,
	title = {Tree-{Structured} {Parzen} {Estimator}: {Understanding} {Its} {Algorithm} {Components} and {Their} {Roles} for {Better} {Empirical} {Performance}},
	shorttitle = {Tree-{Structured} {Parzen} {Estimator}},
	url = {http://arxiv.org/abs/2304.11127},
	doi = {10.48550/arXiv.2304.11127},
	urldate = {2026-01-26},
	publisher = {arXiv},
	author = {Watanabe, Shuhei},
	month = sep,
	year = {2025},
	note = {arXiv:2304.11127 [cs]},
}


%% file: ryan-cites-long.bib
@inproceedings{pillars,
	address = {Philadelphia, PA, USA},
	series = {{MAPL} 2018},
	title = {The three pillars of machine programming},
	isbn = {978-1-4503-5834-7},
	url = {https://doi.org/10.1145/3211346.3211355},
	doi = {10.1145/3211346.3211355},
	urldate = {2020-02-13},
	booktitle = {Proceedings of the 2nd {ACM} {SIGPLAN} {International} {Workshop} on {Machine} {Learning} and {Programming} {Languages}},
	publisher = {Association for Computing Machinery},
	author = {Gottschlich, Justin and Solar-Lezama, Armando and Tatbul, Nesime and Carbin, Michael and Rinard, Martin and Barzilay, Regina and Amarasinghe, Saman and Tenenbaum, Joshua B. and Mattson, Tim},
	month = jun,
	year = {2018},
	pages = {69--80},
}

@misc{bandit_survey,
	title = {A {Survey} on {Contextual} {Multi}-armed {Bandits}},
	url = {http://arxiv.org/abs/1508.03326},
	urldate = {2020-01-30},
	publisher = {arXiv},
	author = {Zhou, Li},
	month = feb,
	year = {2016},
}

@article{deep_cmab,
	series = {{arXiv} '18},
	title = {Deep {Contextual} {Multi}-armed {Bandits}},
	url = {http://arxiv.org/abs/1807.09809},
	urldate = {2020-01-30},
	journal = {arXiv:1807.09809 [cs, stat]},
	author = {Collier, Mark and Llorens, Hector Urdiales},
	month = jul,
	year = {2018},
}

@inproceedings{flood,
	series = {{MLForSystems} @ {NeurIPS} '19},
	title = {Learning {Multi}-dimensional {Indexing}},
	copyright = {All rights reserved},
	booktitle = {{ML} for {Systems} at {NeurIPS}},
	author = {Nathan, Vikram and Ding, Jialin and Alizadeh, Mohammad and Kraska, Tim},
	month = dec,
	year = {2019},
}

@inproceedings{deep_card_est2,
	series = {{CIDR} '19},
	title = {Learned {Cardinalities}: {Estimating} {Correlated} {Joins} with {Deep} {Learning}},
	shorttitle = {Learned {Cardinalities}},
	url = {http://arxiv.org/abs/1809.00677},
	urldate = {2018-09-17},
	booktitle = {9th {Biennial} {Conference} on {Innovative} {Data} {Systems} {Research}},
	author = {Kipf, Andreas and Kipf, Thomas and Radke, Bernhard and Leis, Viktor and Boncz, Peter and Kemper, Alfons},
	year = {2019},
}

@inproceedings{eddies,
	address = {New York, NY, USA},
	series = {{SIGMOD} '00},
	title = {Eddies: {Continuously} {Adaptive} {Query} {Processing}},
	isbn = {978-1-58113-217-5},
	shorttitle = {Eddies},
	url = {http://doi.acm.org/10.1145/342009.335420},
	doi = {10.1145/342009.335420},
	urldate = {2019-05-27},
	booktitle = {Proceedings of the 2000 {ACM} {SIGMOD} {International} {Conference} on {Management} of {Data}},
	publisher = {ACM},
	author = {Avnur, Ron and Hellerstein, Joseph M.},
	year = {2000},
	note = {event-place: Dallas, Texas, USA},
	pages = {261--272},
}

@article{howgood,
	series = {{VLDB} '15},
	title = {How {Good} {Are} {Query} {Optimizers}, {Really}?},
	volume = {9},
	issn = {2150-8097},
	url = {http://dx.doi.org/10.14778/2850583.2850594},
	doi = {10.14778/2850583.2850594},
	number = {3},
	urldate = {2018-02-27},
	journal = {PVLDB},
	author = {Leis, Viktor and Gubichev, Andrey and Mirchev, Atanas and Boncz, Peter and Kemper, Alfons and Neumann, Thomas},
	year = {2015},
	pages = {204--215},
}

@article{q,
	series = {Machine learning '92},
	title = {Q-learning},
	volume = {8},
	number = {3-4},
	journal = {Machine learning},
	author = {Watkins, Christopher JCH and Dayan, Peter},
	year = {1992},
	pages = {279--292},
}

@inproceedings{adaptive_science,
	series = {{ICDE} '08},
	title = {Adaptive {Segmentation} for {Scientific} {Databases}},
	doi = {10.1109/ICDE.2008.4497573},
	booktitle = {Proceedings of the 24th {International} {Conference} on {Data} {Engineering}},
	author = {Ivanova, M. and Kersten, M. L. and Nes, N.},
	month = apr,
	year = {2008},
	pages = {1412--1414},
}

@book{cart,
	address = {Boca Raton},
	title = {Classification and regression trees},
	isbn = {978-0-412-04841-8},
	language = {eng},
	publisher = {Chapman \& Hall},
	author = {Breiman, Leo and Friedman, Jerome and Olshen, Richard and Stone, Charles},
	year = {1998},
}

@article{skinnerdb,
	series = {{VLDB} '18},
	title = {{SkinnerDB}: {Regret}-bounded {Query} {Evaluation} via {Reinforcement} {Learning}},
	volume = {11},
	issn = {2150-8097},
	shorttitle = {{SkinnerDB}},
	url = {https://doi.org/10.14778/3229863.3236263},
	doi = {10.14778/3229863.3236263},
	number = {12},
	urldate = {2019-01-31},
	journal = {PVLDB},
	author = {Trummer, Immanuel and Moseley, Samuel and Maram, Deepak and Jo, Saehan and Antonakakis, Joseph},
	year = {2018},
	pages = {2074--2077},
}

@inproceedings{sql_embed,
	series = {{CIDR} '19},
	title = {Database-{Agnostic} {Workload} {Management}},
	booktitle = {9th {Biennial} {Conference} on {Innovative} {Data} {Systems} {Research}},
	author = {{Shrainik Jain} and {Jiaqi Yan} and {Thiery Cruanes} and {Bill Howe}},
	year = {2019},
}

@book{rl_book,
	address = {Cambridge, MA, USA},
	edition = {1st},
	series = {{MIT} {Press} '98},
	title = {Introduction to {Reinforcement} {Learning}},
	isbn = {978-0-262-19398-6},
	publisher = {MIT Press},
	author = {Sutton, Richard S. and Barto, Andrew G.},
	year = {1998},
}

@article{adaptive_qp_rl,
	series = {Technical {Report}, 08},
	title = {A {Reinforcement} {Learning} {Approach} for {Adaptive} {Query} {Processing}},
	journal = {Technical Reports},
	author = {Tzoumas, Kostas and Sellis, Timos and Jensen, Christian},
	month = jun,
	year = {2008},
}

@article{ppo,
	series = {{arXiv} '17},
	title = {Proximal {Policy} {Optimization} {Algorithms}},
	url = {http://arxiv.org/abs/1707.06347},
	urldate = {2018-02-27},
	journal = {arXiv:1707.06347 [cs]},
	author = {Schulman, John and Wolski, Filip and Dhariwal, Prafulla and Radford, Alec and Klimov, Oleg},
	month = jul,
	year = {2017},
	note = {arXiv: 1707.06347},
}

@article{dqn,
	series = {Nature '15},
	title = {Human-level control through deep reinforcement learning},
	volume = {518},
	number = {7540},
	journal = {Nature},
	author = {Mnih, Volodymyr and Kavukcuoglu, Koray and Silver, David and Rusu, Andrei A. and Veness, Joel and Bellemare, Marc G. and Graves, Alex and Riedmiller, Martin and Fidjeland, Andreas K. and Ostrovski, Georg},
	year = {2015},
	pages = {529--533},
}

@inproceedings{reinforce,
	series = {Machine {Learning} '92},
	title = {Simple statistical gradient-following algorithms for connectionist reinforcement learning},
	booktitle = {Machine {Learning}},
	author = {Williams, Ronald J.},
	year = {1992},
	pages = {229--256},
}

@article{wisedb-vldb,
	series = {{VLDB} '16},
	title = {{WiSeDB}: {A} {Learning}-based {Workload} {Management} {Advisor} for {Cloud} {Databases}},
	volume = {9},
	copyright = {All rights reserved},
	issn = {2150-8097},
	url = {http://dx.doi.org/10.14778/2977797.2977804},
	doi = {10.14778/2977797.2977804},
	number = {10},
	journal = {PVLDB},
	author = {Marcus, Ryan and Papaemmanouil, Olga},
	year = {2016},
	note = {tex.acmid= 2977804
tex.issue\_date= June 2016
tex.numpages= 12},
	pages = {780--791},
}

@article{cuttlefish,
	series = {{arXiv} '18},
	title = {Cuttlefish: {A} {Lightweight} {Primitive} for {Adaptive} {Query} {Processing}},
	shorttitle = {Cuttlefish},
	url = {https://arxiv.org/abs/1802.09180},
	language = {en},
	urldate = {2018-07-19},
	journal = {arXiv preprint},
	author = {Kaftan, Tomer and Balazinska, Magdalena and Cheung, Alvin and Gehrke, Johannes},
	month = feb,
	year = {2018},
}

@inproceedings{tpcds,
	address = {Seoul, Korea},
	series = {{VLDB} '06},
	title = {The {Making} of {TPC}-{DS}},
	url = {http://dl.acm.org/citation.cfm?id=1182635.1164217},
	urldate = {2018-09-16},
	booktitle = {{VLDB}},
	publisher = {VLDB Endowment},
	author = {Nambiar, Raghunath Othayoth and Poess, Meikel},
	year = {2006},
	pages = {1049--1058},
}

@inproceedings{deep_entity,
	address = {New York, NY, USA},
	series = {{SIGMOD} '18},
	title = {Deep {Learning} for {Entity} {Matching}: {A} {Design} {Space} {Exploration}},
	isbn = {978-1-4503-4703-7},
	shorttitle = {Deep {Learning} for {Entity} {Matching}},
	url = {http://doi.acm.org/10.1145/3183713.3196926},
	doi = {10.1145/3183713.3196926},
	urldate = {2018-08-21},
	booktitle = {Proceedings of the 2018 {International} {Conference} on {Management} of {Data}},
	publisher = {ACM},
	author = {Mudgal, Sidharth and Li, Han and Rekatsinas, Theodoros and Doan, AnHai and Park, Youngchoon and Krishnan, Ganesh and Deep, Rohit and Arcaute, Esteban and Raghavendra, Vijay},
	year = {2018},
	pages = {19--34},
}

@misc{url-postgres,
	title = {{PostgreSQL} database, http://www.postgresql.org/},
	url = {http://www.postgresql.org/},
	year = {2024},
}

@article{fastgres,
	series = {{VLDB} '23},
	title = {{FASTgres}: {Making} {Learned} {Query} {Optimizer} {Hinting} {Effective}},
	volume = {16},
	issn = {2150-8097},
	shorttitle = {{FASTgres}},
	url = {https://dl.acm.org/doi/10.14778/3611479.3611528},
	doi = {10.14778/3611479.3611528},
	number = {11},
	urldate = {2023-09-29},
	journal = {Proceedings of the VLDB Endowment},
	author = {Woltmann, Lucas and Thiessat, Jerome and Hartmann, Claudio and Habich, Dirk and Lehner, Wolfgang},
	month = aug,
	year = {2023},
	pages = {3310--3322},
}

@article{neo,
	series = {{VLDB} '19},
	title = {Neo: {A} {Learned} {Query} {Optimizer}},
	volume = {12},
	copyright = {All rights reserved},
	issn = {2150-8097},
	doi = {10.14778/3342263.3342644},
	number = {11},
	journal = {PVLDB},
	author = {Marcus, Ryan and Negi, Parimarjan and Mao, Hongzi and Zhang, Chi and Alizadeh, Mohammad and Kraska, Tim and Papaemmanouil, Olga and Tatbul, Nesime},
	year = {2019},
	pages = {1705--1718},
}

@inproceedings{auto-wlm,
	series = {{SIGMOD} '23},
	title = {Auto-{WLM}: {Machine} {Learning} {Enhanced} {Workload} {Management} in {Amazon} {Redshift}},
	copyright = {All rights reserved},
	doi = {https://doi.org/10.1145/3555041.3589677},
	booktitle = {Companion of the 2023 {International} {Conference} on {Management} of {Data}},
	author = {Saxena, Gaurav and Rahman, Mohammad and Chainani, Naresh and Lin, Chunbin and Caragea, George and Chowdhury, Fahim and Marcus, Ryan and Kraska, Tim and Pandis, Ippokratis and Narayanaswamy, Balakrishnan (Murali)},
	month = jun,
	year = {2023},
}

@inproceedings{qo_rank,
	series = {{BTW} '23},
	title = {Learn {What} {Really} {Matters}: {A} {Learning}-to-{Rank} {Approach} for {ML}-based {Query} {Optimization}},
	doi = {10.18420/BTW2023-25},
	booktitle = {Database {Systems} for {Business}, {Technology}, and the {Web} 2023},
	publisher = {Gesellschaft für Informatik e.V.},
	author = {Behr, Henriette and Markl, Volker and Kaoudi, Zoi},
	editor = {König-Ries, Birgitta and Scherzinger, Stefanie and Lehner, Wolfgang and Vossen, Gottfried},
	year = {2023},
}

@article{tsunami,
	title = {Tsunami: a learned multi-dimensional index for correlated data and skewed workloads},
	volume = {14},
	issn = {2150-8097},
	shorttitle = {Tsunami},
	url = {https://dl.acm.org/doi/10.14778/3425879.3425880},
	doi = {10.14778/3425879.3425880},
	language = {en},
	number = {2},
	urldate = {2023-02-28},
	journal = {Proceedings of the VLDB Endowment},
	author = {Ding, Jialin and Nathan, Vikram and Alizadeh, Mohammad and Kraska, Tim},
	month = oct,
	year = {2020},
	pages = {74--86},
}

@article{learned_xaction,
	series = {{VLDB} '11},
	title = {On {Predictive} {Modeling} for {Optimizing} {Transaction} {Execution} in {Parallel} {OLTP} {Systems}},
	volume = {5},
	doi = {10.14778/2078324.2078325},
	number = {2},
	journal = {PVLDB},
	author = {Pavlo, Andrew and P. C. Jones, Evan and Zdonik, Stan},
	year = {2011},
	pages = {86--96},
}

@inproceedings{buffer_sched,
	address = {Tokyo, Japan},
	series = {{AIDB}@{VLDB} '20},
	title = {Buffer {Pool} {Aware} {Query} {Scheduling} via {Deep} {Reinforcement} {Learning}},
	copyright = {All rights reserved},
	url = {https://drive.google.com/file/d/1trNYAcQ3S71SHu5dbtkBR2hjcK-VWFSx/view?usp=sharing},
	booktitle = {2nd {International} {Workshop} on {Applied} {AI} for {Database} {Systems} and {Applications}},
	author = {Zhang, Chi and Marcus, Ryan and Kleiman, Anat and Papaemmanouil, Olga},
	editor = {He, Bingsheng and Reinwald, Berthold and Wu, Yingjun},
	year = {2020},
}

@inproceedings{lsched,
	address = {New York, NY, USA},
	series = {{SIGMOD} '22},
	title = {{LSched}: {A} {Workload}-{Aware} {Learned} {Query} {Scheduler} for {Analytical} {Database} {Systems}},
	isbn = {978-1-4503-9249-5},
	shorttitle = {{LSched}},
	url = {https://doi.org/10.1145/3514221.3526158},
	doi = {10.1145/3514221.3526158},
	urldate = {2023-02-28},
	booktitle = {Proceedings of the 2022 {International} {Conference} on {Management} of {Data}},
	publisher = {Association for Computing Machinery},
	author = {Sabek, Ibrahim and Ukyab, Tenzin Samten and Kraska, Tim},
	month = jun,
	year = {2022},
	pages = {1228--1242},
}

@inproceedings{balsa,
	address = {New York, NY, USA},
	series = {{SIGMOD} '22},
	title = {Balsa: {Learning} a {Query} {Optimizer} {Without} {Expert} {Demonstrations}},
	isbn = {978-1-4503-9249-5},
	shorttitle = {Balsa},
	url = {https://doi.org/10.1145/3514221.3517885},
	doi = {10.1145/3514221.3517885},
	urldate = {2022-09-10},
	booktitle = {Proceedings of the 2022 {International} {Conference} on {Management} of {Data}},
	publisher = {Association for Computing Machinery},
	author = {Yang, Zongheng and Chiang, Wei-Lin and Luan, Sifei and Mittal, Gautam and Luo, Michael and Stoica, Ion},
	month = jun,
	year = {2022},
	pages = {931--944},
}

@inproceedings{bao,
	address = {China},
	series = {{SIGMOD} '21},
	title = {Bao: {Making} {Learned} {Query} {Optimization} {Practical}},
	copyright = {All rights reserved},
	isbn = {978-1-4503-8343-1},
	doi = {10.1145/3448016.3452838},
	booktitle = {Proceedings of the 2021 {International} {Conference} on {Management} of {Data}},
	author = {Marcus, Ryan and Negi, Parimarjan and Mao, Hongzi and Tatbul, Nesime and Alizadeh, Mohammad and Kraska, Tim},
	month = jun,
	year = {2021},
	note = {Award: 'best paper award'},
}

@inproceedings{ml_index,
	address = {New York, NY, USA},
	series = {{SIGMOD} '18},
	title = {The {Case} for {Learned} {Index} {Structures}},
	isbn = {978-1-4503-4703-7},
	shorttitle = {Learned {Index} {Structures}},
	url = {http://doi.acm.org/10.1145/3183713.3196909},
	doi = {10.1145/3183713.3196909},
	urldate = {2018-08-21},
	booktitle = {Proceedings of the 2018 {International} {Conference} on {Management} of {Data}},
	publisher = {ACM},
	author = {Kraska, Tim and Beutel, Alex and Chi, Ed H. and Dean, Jeffrey and Polyzotis, Neoklis},
	year = {2018},
}

@inproceedings{lstm_jo,
	series = {{ICDE} '20},
	title = {Reinforcement {Learning} with {Tree}-{LSTM} for {Join} {Order} {Selection}},
	doi = {10.1109/ICDE48307.2020.00116},
	booktitle = {2020 {IEEE} 36th {International} {Conference} on {Data} {Engineering}},
	author = {Yu, Xiang and Li, Guoliang and Chai, Chengliang and Tang, Nan},
	month = apr,
	year = {2020},
	note = {ISSN: 2375-026X},
	pages = {1297--1308},
}

@inproceedings{rejoin_tail,
	address = {Portland, Oregon},
	series = {{aiDM} '20},
	title = {Research challenges in deep reinforcement learning-based join query optimization},
	isbn = {978-1-4503-8029-4},
	url = {https://doi.org/10.1145/3401071.3401657},
	doi = {10.1145/3401071.3401657},
	urldate = {2020-07-04},
	booktitle = {Proceedings of the {Third} {International} {Workshop} on {Exploiting} {Artificial} {Intelligence} {Techniques} for {Data} {Management}},
	publisher = {Association for Computing Machinery},
	author = {Guo, Runsheng Benson and Daudjee, Khuzaima},
	month = jun,
	year = {2020},
	pages = {1--6},
}

@article{lero,
	series = {{VLDB} '23},
	title = {Lero: {A} {Learning}-to-{Rank} {Query} {Optimizer}},
	volume = {16},
	issn = {2150-8097},
	shorttitle = {Lero},
	url = {https://doi.org/10.14778/3583140.3583160},
	doi = {10.14778/3583140.3583160},
	number = {6},
	urldate = {2023-12-05},
	journal = {Proceedings of the VLDB Endowment},
	author = {Zhu, Rong and Chen, Wei and Ding, Bolin and Chen, Xingguang and Pfadler, Andreas and Wu, Ziniu and Zhou, Jingren},
	month = feb,
	year = {2023},
	pages = {1466--1479},
}

@inproceedings{duckdb,
	address = {New York, NY, USA},
	series = {{SIGMOD} '19},
	title = {{DuckDB}: an {Embeddable} {Analytical} {Database}},
	isbn = {978-1-4503-5643-5},
	shorttitle = {{DuckDB}},
	url = {https://dl.acm.org/doi/10.1145/3299869.3320212},
	doi = {10.1145/3299869.3320212},
	urldate = {2024-01-14},
	booktitle = {Proceedings of the 2019 {International} {Conference} on {Management} of {Data}},
	publisher = {Association for Computing Machinery},
	author = {Raasveldt, Mark and Mühleisen, Hannes},
	month = jun,
	year = {2019},
	pages = {1981--1984},
}

@article{roq,
	title = {Roq: {Robust} {Query} {Optimization} {Based} on a {Risk}-aware {Learned} {Cost} {Model}},
	copyright = {arXiv.org perpetual, non-exclusive license},
	shorttitle = {Roq},
	url = {https://arxiv.org/abs/2401.15210},
	doi = {10.48550/ARXIV.2401.15210},
	urldate = {2024-03-21},
	author = {Kamali, Amin and Kantere, Verena and Zuzarte, Calisto and Corvinelli, Vincent},
	year = {2024},
}

@article{eraser_lqo,
	series = {{VLDB} '24},
	title = {Eraser: {Eliminating} {Performance} {Regression}  on {Learned} {Query} {Optimizer}},
	volume = {17},
	doi = {10.14778/3641204.3641205},
	language = {en},
	number = {5},
	journal = {PVLDB},
	author = {Weng, Lianggui and Zhu, Rong and Wu, Di and Ding, Bolin and Zheng, Bolong and Zhou, Jingren},
	year = {2024},
	pages = {926--938},
}

@article{pilotscope,
	series = {{VLDB} '24},
	title = {{PilotScope}: {Steering} {Databases} with {Machine} {Learning} {Drivers}},
	volume = {17},
	doi = {10.14778/3641204.3641209},
	language = {en},
	number = {5},
	journal = {PVLDB},
	author = {Zhu, Rong and Weng, Lianggui and Wei, Wenqing and Wu, Di and Peng, Jiazhen and Wang, Yifan and Ding, Bolin and Lian, Defu and Zheng, Bolong and Zhou, Jingren},
	year = {2024},
	pages = {980--993},
}

@inproceedings{morsel,
	address = {Snowbird Utah USA},
	series = {{SIGMOD} '14},
	title = {Morsel-driven parallelism: a {NUMA}-aware query evaluation framework for the many-core age},
	isbn = {978-1-4503-2376-5},
	shorttitle = {Morsel-driven parallelism},
	url = {https://dl.acm.org/doi/10.1145/2588555.2610507},
	doi = {10.1145/2588555.2610507},
	language = {en},
	urldate = {2024-06-07},
	booktitle = {Proceedings of the 2014 {ACM} {SIGMOD} {International} {Conference} on {Management} of {Data}},
	publisher = {ACM},
	author = {Leis, Viktor and Boncz, Peter and Kemper, Alfons and Neumann, Thomas},
	month = jun,
	year = {2014},
	pages = {743--754},
}

@article{reopt,
	series = {{ICDE} '19},
	title = {How {I} {Learned} to {Stop} {Worrying} and {Love} {Re}-optimization},
	copyright = {https://ieeexplore.ieee.org/Xplorehelp/downloads/license-information/IEEE.html},
	url = {https://ieeexplore.ieee.org/document/8731491/},
	doi = {10.1109/ICDE.2019.00191},
	urldate = {2024-10-16},
	journal = {2019 IEEE 35th International Conference on Data Engineering (ICDE)},
	author = {Perron, Matthew and Shang, Zeyuan and Kraska, Tim and Stonebraker, Michael},
	month = apr,
	year = {2019},
	note = {Conference Name: 2019 IEEE 35th International Conference on Data Engineering (ICDE)
ISBN: 9781538674741
Place: Macao, Macao
Publisher: IEEE},
	pages = {1758--1761},
}

@misc{hero_qo,
	title = {{HERO}: {Hint}-{Based} {Efficient} and {Reliable} {Query} {Optimizer}},
	shorttitle = {{HERO}},
	url = {http://arxiv.org/abs/2412.02372},
	doi = {10.48550/arXiv.2412.02372},
	urldate = {2025-01-24},
	publisher = {arXiv},
	author = {Zinchenko, Sergey and Iazov, Sergey},
	month = dec,
	year = {2024},
	note = {arXiv:2412.02372 [cs]},
}

@article{concurrent_lqo,
	title = {Lemo: {A} {Cache}-{Enhanced} {Learned} {Optimizer} for {Concurrent} {Queries}},
	volume = {1},
	shorttitle = {Lemo},
	url = {https://dl.acm.org/doi/10.1145/3626734},
	doi = {10.1145/3626734},
	number = {4},
	urldate = {2025-01-24},
	journal = {Proc. ACM Manag. Data},
	author = {Mo, Songsong and Chen, Yile and Wang, Hao and Cong, Gao and Bao, Zhifeng},
	month = dec,
	year = {2023},
	pages = {247:1--247:26},
}

@article{workload_reopt,
	series = {{VLDB} '19},
	title = {Guided automated learning for query workload re-optimization},
	volume = {12},
	issn = {2150-8097},
	url = {https://dl.acm.org/doi/10.14778/3352063.3352120},
	doi = {10.14778/3352063.3352120},
	language = {en},
	number = {12},
	urldate = {2025-02-02},
	journal = {Proceedings of the VLDB Endowment},
	author = {Damasio, Guilherme and Corvinelli, Vincent and Godfrey, Parke and Mierzejewski, Piotr and Mihaylov, Alex and Szlichta, Jaroslaw and Zuzarte, Calisto},
	month = aug,
	year = {2019},
	pages = {2010--2021},
}

@inproceedings{rio_reopt,
	address = {New York, NY, USA},
	series = {{SIGMOD} '05},
	title = {Proactive re-optimization with {Rio}},
	isbn = {978-1-59593-060-6},
	url = {https://dl.acm.org/doi/10.1145/1066157.1066294},
	doi = {10.1145/1066157.1066294},
	urldate = {2025-02-07},
	booktitle = {Proceedings of the 2005 {ACM} {SIGMOD} international conference on {Management} of data},
	publisher = {Association for Computing Machinery},
	author = {Babu, Shivnath and Bizarro, Pedro and DeWitt, David},
	month = jun,
	year = {2005},
	pages = {936--938},
}

@inproceedings{inc_reopt,
	address = {New York, NY, USA},
	series = {{SIGMOD} '16},
	title = {Enabling {Incremental} {Query} {Re}-{Optimization}},
	isbn = {978-1-4503-3531-7},
	url = {https://dl.acm.org/doi/10.1145/2882903.2915212},
	doi = {10.1145/2882903.2915212},
	urldate = {2025-02-07},
	booktitle = {Proceedings of the 2016 {International} {Conference} on {Management} of {Data}},
	publisher = {Association for Computing Machinery},
	author = {Liu, Mengmeng and Ives, Zachary G. and Loo, Boon Thau},
	month = jun,
	year = {2016},
	pages = {1705--1720},
}

@article{leon_qo,
	series = {{VLDB} '23},
	title = {{LEON}: {A} {New} {Framework} for {ML}-{Aided} {Query} {Optimization}},
	volume = {16},
	issn = {2150-8097},
	shorttitle = {{LEON}},
	url = {https://dl.acm.org/doi/10.14778/3598581.3598597},
	doi = {10.14778/3598581.3598597},
	number = {9},
	urldate = {2025-02-07},
	journal = {Proc. VLDB Endow.},
	author = {Chen, Xu and Chen, Haitian and Liang, Zibo and Liu, Shuncheng and Wang, Jinghong and Zeng, Kai and Su, Han and Zheng, Kai},
	month = may,
	year = {2023},
	pages = {2261--2273},
}

@article{mlos,
	series = {{VLDB} '24},
	title = {{MLOS} in {Action}: {Bridging} the {Gap} {Between} {Experimentation} and {Auto}-{Tuning} in the {Cloud}},
	volume = {17},
	issn = {2150-8097},
	shorttitle = {{MLOS} in {Action}},
	url = {https://dl.acm.org/doi/10.14778/3685800.3685852},
	doi = {10.14778/3685800.3685852},
	number = {12},
	urldate = {2025-05-08},
	journal = {Proc. VLDB Endow.},
	author = {Kroth, Brian and Matusevych, Sergiy and Alotaibi, Rana and Zhu, Yiwen and Gruenheid, Anja and Tian, Yuanyuan},
	month = aug,
	year = {2024},
	pages = {4269--4272},
}

@misc{smartchoices,
	title = {{SmartChoices}: {Hybridizing} {Programming} and {Machine} {Learning}},
	shorttitle = {{SmartChoices}},
	url = {http://arxiv.org/abs/1810.00619},
	doi = {10.48550/arXiv.1810.00619},
	urldate = {2025-05-08},
	publisher = {arXiv},
	author = {Carbune, Victor and Coppey, Thierry and Daryin, Alexander and Deselaers, Thomas and Sarda, Nikhil and Yagnik, Jay},
	month = jun,
	year = {2019},
	note = {arXiv:1810.00619 [cs]},
}

@article{pgo,
	series = {{SPE} '91},
	title = {Using profile information to assist classic code optimizations},
	volume = {21},
	copyright = {Copyright © 1991 John Wiley \& Sons, Ltd},
	issn = {1097-024X},
	url = {https://onlinelibrary.wiley.com/doi/abs/10.1002/spe.4380211204},
	doi = {10.1002/spe.4380211204},
	language = {en},
	number = {12},
	urldate = {2025-07-06},
	journal = {Software: Practice and Experience},
	author = {Chang, Pohua P. and Mahlke, Scott A. and Hwu, Wen-Mei W.},
	year = {1991},
	note = {\_eprint: https://onlinelibrary.wiley.com/doi/pdf/10.1002/spe.4380211204},
	pages = {1301--1321},
}

@inproceedings{microadapt,
	address = {New York, NY, USA},
	series = {{SIGMOD} '13},
	title = {Micro adaptivity in {Vectorwise}},
	isbn = {978-1-4503-2037-5},
	url = {https://dl.acm.org/doi/10.1145/2463676.2465292},
	doi = {10.1145/2463676.2465292},
	urldate = {2025-07-06},
	booktitle = {Proceedings of the 2013 {ACM} {SIGMOD} {International} {Conference} on {Management} of {Data}},
	publisher = {Association for Computing Machinery},
	author = {Răducanu, Bogdan and Boncz, Peter and Zukowski, Marcin},
	month = jun,
	year = {2013},
	pages = {1231--1242},
}

@article{excalibur,
	series = {{VLDB} '22},
	title = {Excalibur: {A} {Virtual} {Machine} for {Adaptive} {Fine}-grained {JIT}-{Compiled} {Query} {Execution} based on {VOILA}},
	volume = {16},
	issn = {2150-8097},
	shorttitle = {Excalibur},
	url = {https://dl.acm.org/doi/10.14778/3574245.3574266},
	doi = {10.14778/3574245.3574266},
	language = {en},
	number = {4},
	urldate = {2025-07-06},
	journal = {Proceedings of the VLDB Endowment},
	author = {Gubner, Tim and Boncz, Peter},
	month = dec,
	year = {2022},
	pages = {829--841},
}

@inproceedings{hw_adapt,
	address = {Hirsau, Germany},
	series = {{GI} '23},
	title = {Towards a {Future} of {Fully} {Self}-{Optimizing} {Query} {Engines}},
	language = {en},
	booktitle = {34th {GI}-{Workshop} on {Foundations} of {Databases}},
	publisher = {CEUR Workshop Proceedings},
	author = {Blockhaus, Paul and Durand, Gabriel Campero and Broneske, David and Saake, Gunter},
	month = jun,
	year = {2023},
}

@inproceedings{datafusion,
	address = {New York, NY, USA},
	series = {{SIGMOD} '24},
	title = {Apache {Arrow} {DataFusion}: {A} {Fast}, {Embeddable}, {Modular} {Analytic} {Query} {Engine}},
	isbn = {979-8-4007-0422-2},
	shorttitle = {Apache {Arrow} {DataFusion}},
	url = {https://dl.acm.org/doi/10.1145/3626246.3653368},
	doi = {10.1145/3626246.3653368},
	urldate = {2025-07-05},
	booktitle = {Companion of the 2024 {International} {Conference} on {Management} of {Data}},
	publisher = {Association for Computing Machinery},
	author = {Lamb, Andrew and Shen, Yijie and Heres, Daniël and Chakraborty, Jayjeet and Kabak, Mehmet Ozan and Hsieh, Liang-Chi and Sun, Chao},
	month = jun,
	year = {2024},
	pages = {5--17},
}

@article{yannakakis_debunk,
	series = {{SIGMOD} '25},
	title = {Debunking the {Myth} of {Join} {Ordering}: {Toward} {Robust} {SQL} {Analytics}},
	volume = {3},
	shorttitle = {Debunking the {Myth} of {Join} {Ordering}},
	url = {https://dl.acm.org/doi/10.1145/3725283},
	doi = {10.1145/3725283},
	number = {3},
	urldate = {2025-07-07},
	journal = {Proc. ACM Manag. Data},
	author = {Zhao, Junyi and Su, Kai and Yang, Yifei and Yu, Xiangyao and Koutris, Paraschos and Zhang, Huanchen},
	month = jun,
	year = {2025},
	pages = {146:1--146:28},
}

@article{yannakakis_diamond,
	title = {Robust {Join} {Processing} with {Diamond} {Hardened} {Joins}},
	volume = {17},
	issn = {2150-8097},
	url = {https://dl.acm.org/doi/10.14778/3681954.3681995},
	doi = {10.14778/3681954.3681995},
	number = {11},
	urldate = {2025-07-07},
	journal = {Proc. VLDB Endow.},
	author = {Birler, Altan and Kemper, Alfons and Neumann, Thomas},
	month = jul,
	year = {2024},
	pages = {3215--3228},
}

@misc{yannakakis_ml,
	title = {Selective {Use} of {Yannakakis}' {Algorithm} to {Improve} {Query} {Performance}: {Machine} {Learning} to the {Rescue}},
	shorttitle = {Selective {Use} of {Yannakakis}' {Algorithm} to {Improve} {Query} {Performance}},
	url = {http://arxiv.org/abs/2502.20233},
	doi = {10.48550/arXiv.2502.20233},
	urldate = {2025-07-07},
	publisher = {arXiv},
	author = {Böhm, Daniela and Gottlob, Georg and Lanzinger, Matthias and Longo, Davide and Okulmus, Cem and Pichler, Reinhard and Selzer, Alexander},
	month = jun,
	year = {2025},
	note = {arXiv:2502.20233 [cs]},
}

@article{adaptive_vs_ml,
	title = {Simple {Adaptive} {Query} {Processing} vs. {Learned} {Query} {Optimizers}: {Observations} and {Analysis}},
	volume = {16},
	issn = {2150-8097},
	shorttitle = {Simple {Adaptive} {Query} {Processing} vs. {Learned} {Query} {Optimizers}},
	url = {https://dl.acm.org/doi/10.14778/3611479.3611501},
	doi = {10.14778/3611479.3611501},
	language = {en},
	number = {11},
	urldate = {2025-07-07},
	journal = {Proceedings of the VLDB Endowment},
	author = {Zhang, Yunjia and Chronis, Yannis and Patel, Jignesh M. and Rekatsinas, Theodoros},
	month = jul,
	year = {2023},
	pages = {2962--2975},
}

@misc{arrow_compute,
	title = {arrow::compute - {Rust}},
	url = {https://docs.rs/arrow/latest/arrow/compute/index.html},
	urldate = {2025-07-18},
	journal = {Apache arrow},
	author = {{Apache}},
	note = {https://docs.rs/arrow/latest/arrow/compute/index.html},
}

@article{sqlstorm,
	series = {{VLDB} '25},
	title = {{SQLStorm}: {Taking} {Database} {Benchmarking} into the {LLM} {Era}},
	volume = {18},
	doi = {10.14778/3749646.3749683},
	language = {en},
	number = {11},
	journal = {PVLDB},
	author = {Schmidt, Tobias and Leis, Viktor and Boncz, Peter and Neumann, Thomas},
	year = {2025},
	pages = {4144--4157},
}

@article{sefrqo,
	series = {{SIGMOD} '26},
	title = {{SEFRQO}: {A} {Self}-{Evolving} {Fine}-{Tuned} {RAG}-{Based} {Query} {Optimizer}},
	volume = {3},
	copyright = {All rights reserved},
	doi = {https://doi.org/10.1145/3769826},
	number = {6},
	journal = {Proceedings of the ACM on Management of Data},
	author = {Liu, Hanwen and Zhang, Qihan and Marcus, Ryan and Sabek, Ibrahim},
	month = may,
	year = {2026},
}

@article{databricks_adaptive,
	series = {{VLDB} '24},
	title = {Adaptive and {Robust} {Query} {Execution} for {Lakehouses} at {Scale}},
	volume = {17},
	issn = {2150-8097},
	url = {https://dl.acm.org/doi/10.14778/3685800.3685818},
	doi = {10.14778/3685800.3685818},
	language = {en},
	number = {12},
	urldate = {2026-01-24},
	journal = {Proceedings of the VLDB Endowment},
	author = {Xue, Maryann and Bu, Yingyi and Somani, Abhishek and Fan, Wenchen and Liu, Ziqi and Chen, Steven and Van Hovell, Herman and Samwel, Bart and Mokhtar, Mostafa and Korlapati, Rk and Lam, Andy and Ma, Yunxiao and Ercegovac, Vuk and Li, Jiexing and Behm, Alexander and Li, Yuanjian and Li, Xiao and Krishnamurthy, Sriram and Shukla, Amit and Petropoulos, Michalis and Paranjpye, Sameer and Xin, Reynold and Zaharia, Matei},
	month = aug,
	year = {2024},
	pages = {3947--3959},
}

@inproceedings{cracking,
	address = {Asilomar, California},
	series = {{CIDR} '07},
	title = {Database {Cracking}},
	author = {Idreos, Stratos and Kersten, Martin L. and Manegold, Stefan},
	month = jan,
	year = {2007},
}

@misc{arrow_filter_strategy,
	title = {Arrow filter logic},
	url = {https://github.com/apache/arrow-rs/blob/5695bb33746735dde01b80ee9e7e6bcccbd7bdd4/arrow-select/src/filter.rs#L356},
	language = {en},
	urldate = {2026-01-31},
	journal = {GitHub},
	month = jan,
	year = {2026},
}

@book{effective_sample_size,
	address = {New York},
	title = {Survey {Sampling}},
	isbn = {0-471-10949-5},
	publisher = {John Wiley \& Sons, Inc.},
	author = {Kish, Leslie},
	year = {1965},
}

@article{hawk_adaptive,
	series = {{VLDBJ} '18},
	title = {Generating custom code for efficient query execution on heterogeneous processors},
	volume = {27},
	issn = {1066-8888},
	url = {https://dl.acm.org/doi/10.1007/s00778-018-0512-y},
	doi = {10.1007/s00778-018-0512-y},
	number = {6},
	urldate = {2026-01-31},
	journal = {The VLDB Journal},
	author = {Breβ, Sebastian and Köcher, Bastian and Funke, Henning and Zeuch, Steffen and Rabl, Tilmann and Markl, Volker},
	month = dec,
	year = {2018},
	pages = {797--822},
}

@article{voila,
	title = {Charting the design space of query execution using {VOILA}},
	volume = {14},
	issn = {2150-8097},
	url = {https://dl.acm.org/doi/10.14778/3447689.3447709},
	doi = {10.14778/3447689.3447709},
	number = {6},
	urldate = {2026-01-31},
	journal = {Proc. VLDB Endow.},
	author = {Gubner, Tim and Boncz, Peter},
	month = feb,
	year = {2021},
	pages = {1067--1079},
}

@techreport{cake_appendix,
	type = {Appendix},
	title = {Appendix to {Piece} of {CAKE}: {Adaptive} {Execution} {Engines} via {Microsecond}-{Scale} {Learning}},
	url = {https://rm.cab/cake_appendix},
	institution = {University of Pennsylvania},
	author = {Zhao, Zijie and Marcus, Ryan},
	year = {2026},
}

@inproceedings{rl_conv1,
	address = {Online},
	series = {{NeurIPS} '21},
	title = {Deep {Reinforcement} {Learning} at the {Edge} of the {Statistical} {Precipice}},
	language = {en},
	author = {Agarwal, Rishabh and Schwarzer, Max and Castro, Pablo Samuel and Courville, Aaron and Bellemare, Marc G},
	year = {2021},
}

@article{rl_conv3,
	series = {{JMLR} '24},
	title = {Empirical {Design} in {Reinforcement} {Learning}},
	volume = {25},
	language = {en},
	journal = {Journal of Machine Learning Research},
	author = {Patterson, Andrew and Neumann, Samuel and White, Martha and White, Adam},
	year = {2024},
	pages = {1--63},
}

@inproceedings{rl_conv4,
	address = {Vienna, Austria},
	series = {{ICML} '20},
	title = {Evaluating the {Performance} of {Reinforcement} {Learning} {Algorithms}},
	language = {en},
	booktitle = {Proceedings of the 37 th {International} {Conference} on {Machine} {Learning}},
	author = {Jordan, Scott M and Chandak, Yash and Cohen, Daniel and Zhang, Mengxue and Thomas, Philip S},
	year = {2020},
}

@inproceedings{rl_conv2,
	address = {Hawaii, USA},
	series = {{ICML} '23},
	title = {Revisiting {Bellman} {Errors} for {Offline} {Model} {Selection}},
	language = {en},
	booktitle = {Proceedings of the 40th {International} {Conference} on {Machine} {Learning}},
	author = {Zitovsky, Joshua P},
	year = {2023},
}
